\newif\ifanonymous
\anonymousfalse   %

\ifanonymous
\documentclass[manuscript,screen,anonymous,review]{acmart}
\else
\documentclass[manuscript,screen]{acmart}
\fi

\AtBeginDocument{%
  }

\usepackage{xspace}
\usepackage{xcolor}
\usepackage[most]{tcolorbox}
\usepackage{fontawesome}
\usepackage{framed}
\usepackage{changepage}
\usepackage{multicol}
\usepackage{multirow}
\usepackage{pgffor}
\usepackage[table]{xcolor}
\usepackage{ifthen}

\begin{document}

\definecolor{PAblue}{RGB}{0,122,204}%
\definecolor{PAlightblue}{RGB}{235, 247, 255}%
\definecolor{myred}{RGB}{235,101,93}
\definecolor{mypink}{RGB}{251,233,234}

\newcommand{\tool}{{AgentUXlab\xspace}}
\newcommand{\todo}[1]{{\color{orange} \bfseries TODO: #1}}

\newcommand{\added}[1]{{\color{blue}#1}}

\newcommand{\pBlockQuo}[2]{
\begin{quote}
\emph{``#1''} (US-P#2)
\end{quote}
}

\newcommand{\numFirstStudyParticipants}{12\xspace}
\newcommand{\numSecondStudyParticipants}{14\xspace}

 \newtcbox{\ilabel}[1][]{enhanced,
 box align=base,
 nobeforeafter,
 colback=PAblue,
 colframe=PAblue,
 size=small,
 fontupper=\color{white}\scriptsize\bf\sffamily,
 left=0.2pt,
 right=0.2pt,
 top=0.2pt,
 bottom=0.2pt,
 boxsep=2pt,
 arc=4.5pt,
 #1}

\newcommand{\epQuote}[2]{{\emph{``#1''} (RE-P#2)}}
\newcommand{\pQuote}[2]{{\emph{``#1''} (US-P#2)}}
\newcommand{\actionTheme}[1]{{\faPlayCircleO\xspace\emph{#1}}}
\newcommand{\planTheme}[1]{{\faListOl\xspace\emph{#1}}}
\newcommand{\systemLevelTheme}[1]{{\faGears\xspace\emph{#1}}}
\newcommand{\userInputTheme}[1]{{\faCommentsO\xspace\emph{#1}}}
\newcommand{\conditionsTheme}[1]{{\faExclamationTriangle\xspace\emph{#1}}}

\newtcbox{\formativeCode}[1][]{enhanced,
 box align=base,
 nobeforeafter,
 colback=PAlightblue,
 colframe=PAlightblue,
 fontupper=\small\ttfamily,
 left=0.2pt,
 right=0.2pt,
 top=0.2pt,
 bottom=0.2pt,
 boxsep=0.4pt,
 #1}

\newtcbox{\taxonomyCode}[1][]{enhanced,
 box align=base,
 nobeforeafter,
 colback=mypink,
 colframe=mypink,
 fontupper=\small\ttfamily,
 left=0.2pt,
 right=0.2pt,
 top=0.2pt,
 bottom=0.2pt,
 boxsep=0.4pt,
 #1}

 \newcommand{\scenario}[1]{ 
	\def\FrameCommand{%
		\hspace{0pt}%
		{\color{PAblue}\vrule width 2pt}%
		{\color{white}\vrule width 2pt}%
		\colorbox{white}
	}%
	\MakeFramed{\advance\hsize-\width\FrameRestore}%
	\noindent\hspace{-4.55pt}%
	\begin{adjustwidth}{}{0pt}
		\emph{#1}
		\vspace{-3pt}
	\end{adjustwidth}\endMakeFramed%
}

\NewDocumentCommand{\numSpaces}{m}{%
  \foreach \i in {1,...,#1} {%
    \xspace
  }%
}

\definecolor{mygray2}{RGB}{224, 224, 224}%

\definecolor{boxcolor}{RGB}{238, 223, 204} %
\DeclareRobustCommand{\mybox}[2][gray!20]{%
\begin{tcolorbox}[   %
        breakable,
        left=0pt,
        right=0pt,
        top=0pt,
        bottom=0pt,
        colback=#1,
        colframe=black,
        width=\dimexpr\columnwidth\relax, 
        enlarge left by=0mm,
        boxsep=5pt,
        outer arc=4pt,
        boxrule=.5mm
        ]
        #2
\end{tcolorbox}
}

\newtcolorbox{resultbox}[1][]{
  enhanced,
  colback=gray!5,
  colframe=black!75,
  boxrule=0.8pt,
  arc=2mm,
  left=2mm,
  right=2mm,
  top=1.5mm,
  bottom=1.5mm,
  fonttitle=\bfseries,
  title=Key Result,
  #1
}
\newcommand{\icon}[1]{{\includegraphics[height=1.5\fontcharht\font`\B]{#1}}\xspace}
\newcommand{\meiicon}{\icon{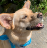}}

\newcommand{\activityOne}[1]{%
\ifnum#1=1
A1. Design the scope and boundaries of the agent%
\else
Design the scope and boundaries of the agent (\textbf{A1})%
\fi
}

\newcommand{\activityTwo}[1]{%
\ifnum#1=1
A2. Design the agent’s information display%
\else
Design the agent’s information display (\textbf{A2})%
\fi
}

\newcommand{\activityThree}[1]{%
\ifnum#1=1
A3. Design the interactions between the agent and user%
\else
Design the interactions between the agent and user (\textbf{A3})%
\fi
}

\newcommand{\activityFour}[1]{%
\ifnum#1=1
A4. Run the agent prototype%
\else
Run the agent prototype (\textbf{A4})%
\fi
}

\newcommand{\activityFive}[1]{%
\ifnum#1=1
A5. Understand the agent’s runtime behavior%
\else
Understand the agent’s runtime behavior (\textbf{A5})%
\fi
}

\newcommand{\capabilityOne}[1]{%
\ifnum#1=1
C1. Use no-code interfaces%
\else
Use no-code interfaces (\textbf{C1})%
\fi
}

\newcommand{\capabilityTwo}[1]{%
\ifnum#1=1
C2. Help constrain the agent’s task space and knowledge of the user%
\else
Help constrain the agent’s task space and knowledge of the user (\textbf{C2})%
\fi
}

\newcommand{\capabilityThree}[1]{%
\ifnum#1=1
C3. Define the UI of the agent in the chat and the environment%
\else
Define the UI of the agent in the chat and the environment (\textbf{C3})%
\fi
}

\newcommand{\capabilityFour}[1]{%
\ifnum#1=1
C4. Provide components that enable the agent to invoke different user interactions%
\else
Provide components that enable the agent to invoke different user interactions (\textbf{C4})%
\fi
}

\newcommand{\capabilityFive}[1]{%
\ifnum#1=1
C5. Provide an environment to run and control the agent%
\else
Provide an environment to run and control the agent (\textbf{C5})%
\fi
}

\newcommand{\capabilitySix}[1]{%
\ifnum#1=1
C6. Help debug the agent’s runtime behavior%
\else
Help debug the agent’s runtime behavior (\textbf{C6})%
\fi
}

\definecolor{boxcolor}{RGB}{238, 223, 204} %
\DeclareRobustCommand{\mybox}[2][gray!20]{%
\begin{tcolorbox}[   %
        breakable,
        left=0pt,
        right=0pt,
        top=0pt,
        bottom=0pt,
        colback=#1,
        colframe=black,
        width=\dimexpr\columnwidth\relax, 
        enlarge left by=0mm,
        boxsep=5pt,
        outer arc=4pt,
        boxrule=.5mm
        ]
        #2
\end{tcolorbox}
}

\title{Understanding User Experiences of Computer Use Agents: Design Space and Opportunities for Building Agent UX Prototypes
}

\author{Jenny T. Liang}
\authornote{Work completed while at Apple}
\email{jtliang@cs.cmu.edu}
\affiliation{%
  \institution{Carnegie Mellon University}
  \city{Pittsburgh}
  \state{Pennsylvania}
  \country{USA}
}

\author{Titus Barik}
\email{tbarik@apple.com}
\affiliation{%
  \institution{Apple}
  \city{Seattle}
  \state{Washington}
  \country{USA}
}

\author{Jeffrey Nichols}
\email{jwnichols@apple.com}
\affiliation{%
  \institution{Apple}
  \city{San Diego}
  \state{California}
  \country{USA}
}

\author{Eldon Schoop}
\email{eldon@apple.com}
\affiliation{%
  \institution{Apple}
  \city{Seattle}
  \state{Washington}
  \country{USA}
}

\author{Ruijia Cheng}
\email{rcheng23@apple.com}
\affiliation{%
  \institution{Apple}
  \city{Seattle}
  \state{Washington}
  \country{USA}
}

\renewcommand{\shortauthors}{Liang et al.}

\begin{abstract}
Computer use agents (or ``agents'') are generative AI that automates actions within user interfaces from user commands.
Current research focuses on training and evaluating the underlying models, leaving these agents' user experience (UX) understudied.
We conducted two studies to understand the design space of agent UX (\textbf{RQ1}) and the support required to prototype it (\textbf{RQ2}).
First, we develop a taxonomy of design considerations for agent UX, comprising 21 subcategories of UX considerations.
Then, through a requirements elicitation study with 12 participants---including six agent experts---we identify five \emph{Activities} and six \emph{Desired Capabilities} needed in tools prototyping agent UX.
Informed by these results, we created \tool, a design probe that enables developers to design agents with different UX approaches for a website and evaluate those experiences by executing prototypes in a browser.
From a user study with 14 participants, we elucidate tooling insights and derive design implications for agent UX prototyping tools.
\end{abstract}

\received{20 February 2007}
\received[revised]{12 March 2009}
\received[accepted]{5 June 2009}

\maketitle

\begin{figure*}[t!]
\centering
\includegraphics[trim=0 0 475 0, clip, width=\linewidth, keepaspectratio, page=1]{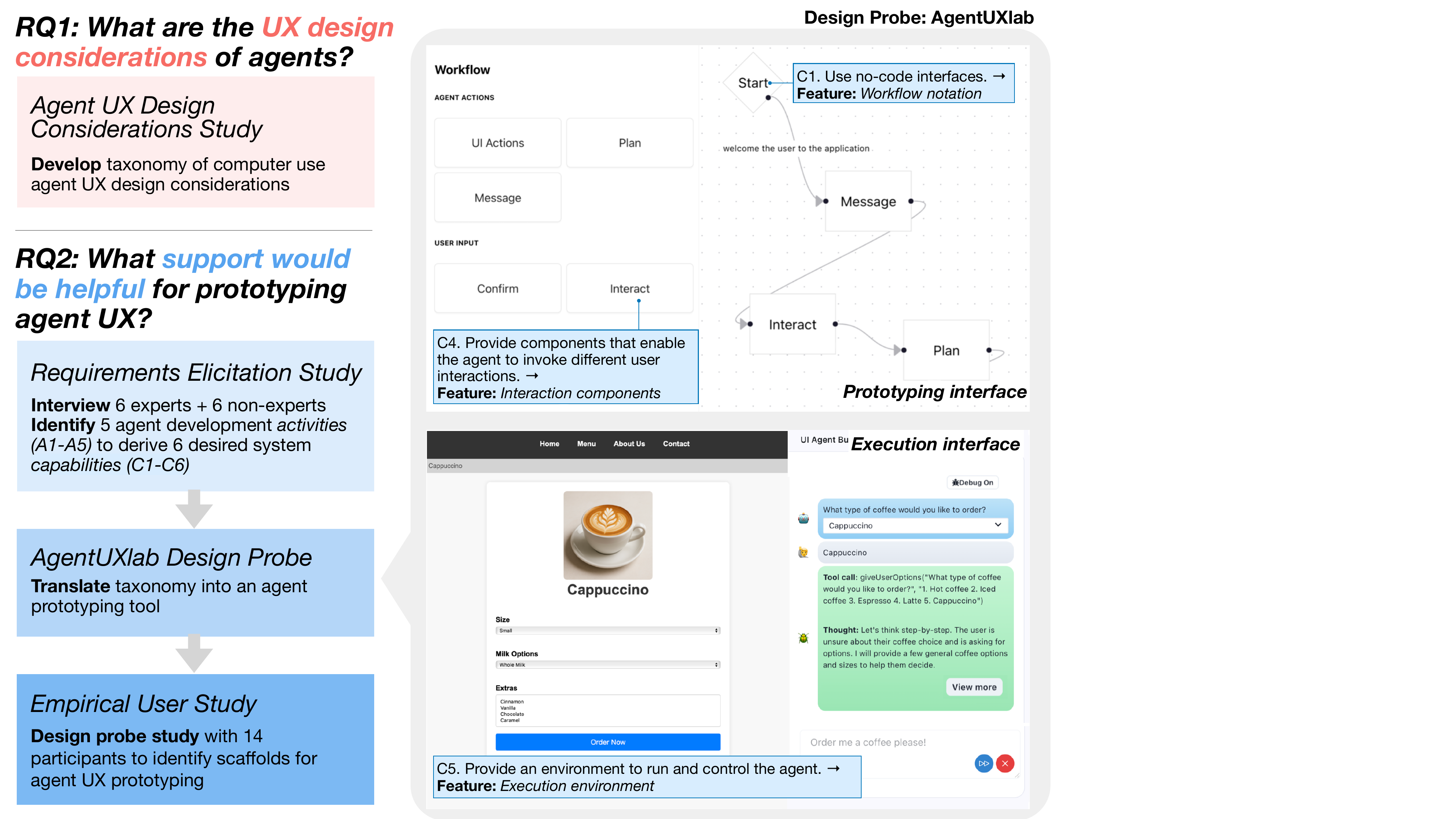} 
\caption{
An overview of the research.
To understand the UX design considerations of computer use agents (\textbf{RQ1}, Section~\ref{sec:taxonomy}), we develop a taxonomy of UX design considerations of agents extracted from existing computer use agents, with feedback from AI and UX practitioners.
To study the support that are needed to support prototyping agent UX (\textbf{RQ2}, Section~\ref{sec:design-probe}), we follow a three step process.
\textbf{Step 1:} We conduct a requirements elicitation study to identify key Activities and Desired Capabilities (such as \textbf{C1}, \textbf{C4}, and \textbf{C5} on the figure) of prototyping systems for agent UX.
\textbf{Step 2:} We develop \tool, a high-fidelity design probe that surfaces the UX design considerations of agent UX and enables prototyping of agent UX.
\textbf{Step 3:} We run an empirical user study to probe how the UX design considerations are explored using \tool.
}
\label{fig:summary-figure}
\end{figure*}

\section{Introduction}
\scenario{Mei is an interaction designer.
She is hired by a coffee shop to prototype a variety of agent UX to understand the best way to have an agent place coffee orders for customers on the website.
As Mei prototypes what the user experience with the agent should be like, she is not sure how proactive the agent should be: whether the agent should frequently ask for user input, perform the task and send a summary at the end, or something in between.
As Mei wants to test these ideas, she wishes there was some kind of scaffold for her to easily prototype these experiences.}

Mei wants to prototype a \emph{computer use agent}---which we refer to in this work as an \emph{agent} for brevity---an AI that automates actions on user interfaces by understanding user habits, preferences, and interests and interacting with available user interface (UI) elements (e.g., scrolling through pages, tapping on buttons) to perform tasks.
In recent years, large language models (LLMs) and multimodal large language models (MLLMs) have rapidly emerged to power computer use agents~\cite{fourney2024magentic, wu2024oscopilot, wang2024mobileagent, he-etal-2024-webvoyager, iong-etal-2024-openwebagent} on desktop~\cite{fourney2024magentic, wu2024oscopilot}, mobile~\cite{wang2024mobileagent}, and web~\cite{he-etal-2024-webvoyager, zhou2024webarena, pan2024webcanvas, zheng-etal-2024-webolympus, iong-etal-2024-openwebagent} environments. 
These have enabled numerous publicly available commercial computer use agents, such as Claude Computer Use Tool~\cite{anthropic2025computer} and OpenAI Operator~\cite{openai2025operator} that perform actions resulting in real-world consequences (e.g., modifying user files, placing online orders)~\cite{zhang2025interaction}.

With the increasing influence of computer use agents, current research has largely focused on training~\cite{he-etal-2024-webvoyager, cai2025large} and evaluating~\cite{pan2024webcanvas, cai2025large, zhou2024webarena, pan2024webcanvas, huq-etal-2025-cowpilot} the models underlying these agents.
However, as illustrated by Mei's experience, an important aspect of agents is designing the user experience (UX)~\cite{horvitz1999principle, shneiderman1997direct, schiaffino2004user, anonymous2026redacted}
\ifanonymous
\footnote{\cite{anonymous2026redacted} is redacted for citing a related work from the authors to adhere in order to TOCHI's author anonymity requirement}
\else
\fi
, which we refer to as \emph{agent UX}.
While the literature offers preliminary guidance on designing agent UX (e.g.,~\cite{horvitz1999principle, shneiderman1997direct, schiaffino2004user, yang2020re}), these recommendations were developed in pre-LLM contexts, whereas modern computer use agents introduce new UX challenges~\cite{bansal2024challenges}.
For example, designing the UX of modern AI requires integrating perspectives from design, machine learning, and engineering~\cite{yang2020re, dove2017ux, yang2018investigating, kayacik2019identifying, subramonyam2025prototyping, yang2025orbit}. 
Prototyping these experiences also requires rapid exploration of prompts and careful consideration of the UI components through which AI agents interact with users~\cite{subramonyam2025prototyping}.

In short, understanding the UX design of agents requires studying \emph{what} the UX design considerations of agents are as well as \emph{how} these considerations can be explored in the design process, as the former can influence the latter~\cite{yang2020re}. 
Given the nascence of computer use agents, we need to understand the underexplored design space of agent UX and what support that individuals involved in designing agent UX need in order to create well-designed agent experiences in the future.

In this work, we aim to reduce these gaps by investigating the following research questions:
\begin{description}
\setlength{\itemsep}{0pt}
\setlength{\parskip}{2pt}
    \item[\textbf{RQ1}] What are the UX design considerations of computer use agents?
    \item[\textbf{RQ2}] 
    What support would be helpful to prototype UX in computer use agents? 
\end{description}

To answer \textbf{RQ1}, we develop a taxonomy of UX design considerations of computer use agents by reviewing existing agents and refining it with interviews from eight UX and AI practitioners.
We then validate the taxonomy with 20 end-users of agents using a Wizard-of-Oz computer use agent.
To explore \textbf{RQ2}, we focus on the process of prototyping as a lens to study designing agent UX, echoing the position of \citet{yang2020re}.
We conducted a three-phase study to investigate how the agent UX design space is explored by translating the taxonomy into a design probe for prototyping agent UX.
To inform the design of the probe, we perform a requirements elicitation study with 12 participants (including six agent experts), where we identify five distinct \emph{Activities} and six \emph{Desired Capabilities} that a system for prototyping computer use agents should afford.
Alongside the taxonomy, we translate the Desired Capabilities into a high-fidelity design probe for agent UX prototyping, known as \tool.
\tool~surfaces agent UX design considerations in a graphical, no-code playground.
It enables developers of agent UX prototypes to design, execute, and iterate upon different agent UX.
We use \tool~not as a deployable tool, but rather an instrument to understand agent UX prototyping.
Therefore, to understand the design process and tooling needs for designing agent UX, using \tool, we run a user study with 14 participants and elicit insights on the support needed for agent prototyping, such as the need for scaffolds that structure the agent UX design space.
Based on these findings, we propose a set of design recommendations for agent UX prototyping systems.

In summary, the contributions of this work are:
\begin{itemize}
    \item \textbf{A taxonomy of the UX design considerations of computer use agents} with four main categories: user queries, explainability of agent activities, user control, and user mental model and expectations (Section~\ref{sec:taxonomy}).
    \item \textbf{A set of design requirements for agent UX prototyping systems derived from a requirements elicitation study}, articulated as five activities and six capabilities that agent UX prototyping systems should support (Section~\ref{sec:formative-study}).
    \item \textbf{\tool, a high-fidelity design probe that instantiates the design requirements and UX design consideration taxonomy}, which has a no-code interface that enables the development of agent prototypes for web-based environments.
    \tool~enables developers to design agents with different UX approaches for a website and evaluate those experiences by executing the prototypes directly in a browser (Section~\ref{sec:tool-design-probe}).
    \item \textbf{Additional insights on the design requirements and support needed to prototype agent UX} based on an empirical user study, with as \tool~as a design probe (Section~\ref{sec:in-situ-study}).
\end{itemize}
This work offers a basis for designing agent UX, with the aim of informing future interfaces to support this activity.

\section{Related Work}

\label{sec:related-work}
Below, we describe related work in the history and state-of-art of computer use agents (Section~\ref{sec:interface-agents}) as well as the importance and challenges in designing user experiences in generative AI systems (Section~\ref{sec:ai-ux-design}). 

\subsection{Computer Use Agents}
\label{sec:interface-agents}

Computer use agents (i.e., AI that performs tasks for users by interacting with UI elements) have been long explored by the HCI community via interface agents that autonomously operate the UI based on a user's command~\cite{lieberman2003agents, lieberman1997autonomous}.
Other researchers have argued against full autonomy of the agent, instead investigating how to combine agents with direct manipulation~\cite{shneiderman1997direct}.
These early works explored a range of applications, such as web browsing~\cite{lieberman1995Letizia}, scheduling~\cite{maes1993learning}, and task automation~\cite{leshed2008coscripter, li2019pumice, li2020multi}.

These early systems lay the foundations for modern computer use agents that can navigate through UIs (e.g., clicking on buttons), powered by recent advancements in (M)LLMs.
Computer use agents can operate in a variety of environments, such as web~\cite{he-etal-2024-webvoyager, huq-etal-2025-cowpilot}, mobile UIs~\cite{wang2024mobileagent, you2024ferret, li2020multi}, and computer operating systems~\cite{wu2024oscopilot} by semantically understanding user input.
Examples of systems include ILuvUI~\cite{jiang2025ILuvUI} and Spotlight~\cite{li2022spotlight}, which support single-screen UI tasks; Ferret-UI~\cite{you2024ferret}, which enables question answering about UIs; WebVoyager~\cite{he-etal-2024-webvoyager}, which automates actions on web browsers based on natural language queries.

Much of the existing research on computer use agents focuses on developing new modeling approaches (e.g.,~\cite{he-etal-2024-webvoyager, cai2025large, you2024ferret}), benchmarks (e.g.,~\cite{zhou2024webarena, pan2024webcanvas, cai2025large}), and evaluation environments (e.g.,~\cite{zhou2024webarena, pan2024webcanvas, huq-etal-2025-cowpilot}) to improve agent performance, leaving the user experience of agents underexplored.
Yet, the HCI community's early investigations on pre-(M)LLM agents demonstrate that agent UX is important.
It yielded design considerations for agents, such as when users should assume control, the UI representation of the agent, what controls an agent should have, and the scope of services~\cite{allen1999mixed, horvitz1999principle, schiaffino2004user, shneiderman1997direct}.
With today's computer use agents, many of these design considerations remain and are even exacerbated, given the increased level of autonomy and generality with the modern agents.
This introduces new questions about designing for trust, transparency, delegation, and control~\cite{bansal2024challenges}.

Therefore, understanding the design space of agent UX in the era of (M)LLM-based computer use agents is essential for designing usable AI.
Yet, this topic remains underexplored, but is slowly garnering interest.
Example work in this space is CowPilot~\cite{huq-etal-2025-cowpilot}, a web agent that allows users to intervene on the agent by providing agent controls.
While these works serve as a helpful starting point to understand agent UX design, currently, there is no comprehensive mapping of the design space for UX of computer use agents, aside from our prior work~\cite{anonymous2026redacted} that this article reports in Section~\ref{sec:taxonomy} and leverages to develop a design probe for agent UX prototyping in Section~\ref{sec:design-probe}.
We address this gap by examining the UX of computer use agents, outlining a taxonomy of UX design considerations for computer use agents.

\subsection{Designing for User Experience in Generative AI}
\label{sec:ai-ux-design}
Considering the UX is crucial in the design of generative AI technologies, yet remains a significant challenge.
\citet{yang2020re} found the UX design challenges of AI at large come from two sources: \emph{capability uncertainty} (i.e., the functionality and potential errors of an AI) and \emph{output complexity} (i.e., the space of what the AI produces as a possible output).

With the UX design of generative AI experiences, similar challenges remain.
Capability uncertainty persists, since stakeholders of generative AI experiences are often unaware or have unrealistic expectations of generative AI capabilities~\cite{zamfirescu2023johnny, dolata2024developmen}.
Further, the output complexity of generative AI is more pronounced, as the output space of generative AI can be any text, image, or more, depending on the model, creating challenges in managing or understanding generative AI outputs~\cite{gero2024supporting, liu2024we}.
In response to these new challenges with generative AI, researchers have proposed design guidelines for generative AI experiences, such as suggesting ideas for prompting and generating multiple outputs~\cite{weisz2024design, subramonyam2024bridging}. 

Crucial to designing AI user experiences is \textit{prototyping}, which requires a range of expertise in design, engineering, and machine learning~\cite{yang2020re, dove2017ux, yang2018investigating, kayacik2019identifying}.
Researchers have long looked at techniques to prototype UX of AI, allowing designers and developers to test existing components representing different aspects of the experience.
For example, early work developed toolkits to develop speech UIs, such as Suede~\cite{klemmer2000suede} and CSLU~\cite{sutton1997cslu}, which enable designers to prototype dialogues by composing pre-defined components that were then tested with a Wizard-Of-Oz protocol.
In the modern generative AI context, tools like PromptInfuser~\cite{petridis2024promptinfuser} allow UX designers of generative AI to prototype both UI and the prompts powering the AI at the same time.

Like other generative AI applications, developing agent UX largely involves writing natural language prompts that are embedded in the application~\cite{zamfirescu2023johnny, liang2025prompts, subramonyam2025prototyping}.
\citet{subramonyam2025prototyping} found that the prototyping process of generative AI UX requires rapid exploration of prompts and consideration of UI components an AI might use to interact with a user.
However, the process of developing agent UX is understudied.
To our knowledge, the most relevant works are agent development tools, including \citet{epperson2025interactive}'s and \citet{hao2025flowforge}'s formative studies on developers of multi-agent workflows.
Documented challenges include wrangling long agent conversations, managing a large design space of how to decompose task, defining agent interactions, assigning agent responsibility, and configuring individual agent behavior.
To assist with the complexities of agent development, several agent developer tools in both research~\cite{hao2025flowforge, epperson2025interactive} and practice~\cite{zapier2026zapier, n8n2026n8n, openai2026agentbuilder, langchain2026langgraph} have emerged. 
Many of these tools, including FlowForge~\cite{hao2025flowforge}, Zapier~\cite{zapier2026zapier}, n8n~\cite{n8n2026n8n}, OpenAI AgentBuilder~\cite{openai2026agentbuilder}, and LangGraph~\cite{langchain2026langgraph}, rely on node-based interfaces to scaffold the process of defining, viewing, and editing agentic workflows.

While the existing literature provides a preliminary understanding of designing agent UX, the important design dimensions of agent UX and methods to support the design of agent UX remain unknown.
We address this gap by developing a taxonomy of agent UX design considerations, and by developing a design probe for agent UX prototyping---\tool---which to our knowledge, is the first system to support this activity.
The design probe surfaces features to explore how to support the design of user interactions, execute agent prototypes, and debug issues in the agent prototype.
By comparison, the literature primarily focuses on how to support the engineering of agentic workflows, rather than building agent UX prototypes.
Inspired by existing approaches that provide users with existing components via node-based builders, we envision combining these with prompt development techniques for agents.
\section{What are the UX design considerations of computer use agents? (RQ1)}
\label{sec:taxonomy}
To understand the UX design space of computer use agents (\textbf{RQ1}), we developed a taxonomy of design considerations for agent UX based on a review of existing computer use agents as well as interviews with eight UX practitioners.
We expect the identified design dimensions to provide a shared vocabulary and useful directions to help designers reason about factors related to the user experience of computer use agents.

\subsection{Methodology}
To produce the taxonomy of UX considerations of computer use agents, we first developed an initial taxonomy by reviewing real-world computer use agents agents, refined it based on interviews with UX and AI practitioners, and then validated it in a user study with a Wizard-of-Oz computer use agent.

\subsubsection*{Protocol}
We elaborate on the development, refinement, and validation phases of the taxonomy below.
All studies involving human participants were approved by our institution's IRB.
We obtained consent from all participants prior to conducting each study session.

\paragraph{Developing the Initial Taxonomy}
We developed the initial taxonomy by reviewing eleven computer use agents that were announced to the public in 2024 to 2025 (see Table \ref{tab:ui_agents}).
These agents represented the state of the art in UI control and operated across a range of platforms, including desktop, browser, and mobile environments.
For each agent, the research team reviewed the demo videos if available, and installed and used the agent for 30 minutes for the ones that were available.
We took field notes during the demo videos and usage sessions, and discussed as a team the significant themes that we observed. 

As a result of those discussions, we developed an initial taxonomy containing a categorization of the design space for LLM-based computer use agents.
The initial taxonomy contained themes and sub-themes of design considerations related to UX.
These were: user prompts, explainability of agent activities, user control, end of task indicators, and task previews.
For each of the themes, we also included examples of design features sourced from the set of computer use agents that we reviewed.
We also included speculative design features that the research team imagined but did not observe among the existing computer use agents.

\begin{table}[h]
\centering
\begin{tabular}{lp{6cm}}
\toprule
\textbf{Platform} & \textbf{Agents} \\
\hline
Desktop & Claude Computer Use Tool \cite{anthropic2025computer}, Adept \cite{adept2025ai}, \newline
OpenAI Operator \cite{openai2025operator}, AIlice \cite{myshell2026ailice}, Magentic-UI \cite{mozannar2025magenticui}, UI-TARS (Computer) \cite{qin2025uitars} \\
\hline
Web Browser & Project Mariner \cite{deepmind2025mariner}, TaxyAI \cite{taxyai2026browser}, 
\newline
AutoGLM (Browser) \cite{liu2024autoglm} \\
\hline
Mobile & UI-TARS (Mobile) \cite{qin2025uitars}, AutoGLM (Mobile) \cite{liu2024autoglm} \\
\bottomrule
\end{tabular}
\caption{Computer use agents reviewed to develop the initial taxonomy.}
\label{tab:ui_agents}
\end{table}

\paragraph{Refining the Taxonomy}
To refine the taxonomy, we presented the initial taxonomy in an interview study with eight practitioners who were designers, engineers, or researchers working in the domains of UX or AI at a large technology company.
The goal of the study was to collect feedback on the initial taxonomy. 
In particular, we wanted to investigate how each theme aligned with design considerations, whether there were any important omissions, and how each theme could be extended with sub-themes (e.g., for the agent mistakes sub-theme, what are the different types of agent mistakes?).  

Each interview study session lasted 45 to 60 minutes.
Participants received \$12 meal voucher.

During each session, a researcher first asked the participant about an example of an LLM-based computer use agent they had used or seen demos of in order to ground their thinking around agents and probe for UX-related considerations.
Example questions included: 
What kind of tasks do you expect a user to achieve with this agent? 
What are some challenges a user might face when interacting with the agent? 

Next, we showed the initial taxonomy to participants via a slide presentation.
We explained each theme and design feature, which was supplemented with screenshots and video snippets from the agents that we reviewed.
We asked for feedback on the taxonomy with the goal of answering the following questions: 

\begin{itemize}
    \item Are there missing themes and sub-themes that are important to the UX of computer use agents? 
    \item How important are each of these themes and sub-themes to the design of computer use agents? 
    \item How can each of the design features potentially impact the user experience? 
\end{itemize}

We recorded the audio and video of each session and took field notes.
We transcribed the recordings and conducted thematic analysis, which we describe in further detail below.
Based on the feedback from participants, we refined the initial taxonomy by re-organizing and renaming the themes, as well as adding new themes, into a final taxonomy.

\paragraph{Validating the Taxonomy}
To validate the finalized taxonomy, we conducted a 90-minute Wizard-of-Oz study sessions~\cite{dahlback1993wizard} with 20 end-users of computer use agents.
Participants of the taxonomy validation study received a \$24 meal voucher.
Participants were presented with a mock computer use agent played by a researcher.
Participants interacted with the agent through a chat interface to complete a series of six tasks for two domains: booking a vacation rental or purchasing items on an e-commerce website.
The tasks were designed to cover a variety of situations, including tasks with ambiguity, errors, and high risks.

During the study session, the participant first discussed their prior experience with computer use agents, as well as any challenges or concerns they had.
Next, we introduced participants to a tutorial of the Wizard-of-Oz agent.
After the tutorial, the participant completed six tasks with the agent through the designated chat interface.
At the end of the session, participants were asked to reflect on their overall user experience with the agent and to suggest any additional features.
Finally, we debriefed participants on the Wizard-of-Oz setup of the agent.

\subsubsection*{Analysis}
To analyze the data gathered from the refinement and validation phases, we followed a provisional coding qualitative analysis approach outlined by~\citet{saldana2009coding}.
The last author of this paper coded all transcripts using the subcategories in the taxonomy and looked for any new subcategories.
In addition, within each subcategory, the last author conducted open coding to iteratively develop themes that extended the example features in the taxonomy and conveyed why users considered particular aspects of UX as important (or unimportant).

For the validation stage only, the last author extracted 612 coded quotes from all participants during the provisional coding process.
In order to validate the coverage of our taxonomy, the first author of the paper coded a sample of 103 quotes (16.8\% of the total validation dataset).
These quotes were randomly sampled from the dataset, with five selected from each subcategory (except the \taxonomyCode{safety} subcategory, where all three quotes were selected).
The first author then independently reviewed the subcategories assigned to the quotes by the last author and marked a set of 19 quotes (18.4\%) that they would
code differently than the original coding.
The two authors then engaged in discussions to resolve disagreements and iterate on the coding, resolving 18 of the 19 quotes whose codes that the authors disagreed on.
Only 1 quote (0.97\%) were decided as being classified incorrectly in the original coding, and was then corrected. 

\subsection{The Taxonomy of UX Design Considerations for Computer Use Agents}
We present the complete taxonomy in Table~\ref{tab:taxonomy}.
The taxonomy follows a three-level hierarchy including \textit{category}, which are the high-level areas of UX considerations in the design space; \taxonomyCode{subcategory}, which are the specific aspects to consider in a particular area; and example features, which are illustrative features supporting the UX consideration that are observed in existing agents or speculated by the research team. 
From the Wizard-of-Oz study, we did not identify any new subcategories, indicating that that all taxonomy subcategories were validated as important aspects of UX considerations during users' interactions with a computer use agent.
In addition to reporting the agent UX design consideration taxonomy, additional details regarding participants' impressions on the specific design considerations are discussed in~\cite{anonymous2026redacted}.

\begin{table*}[]
\small
\begin{tabular}{|l|l|l|}
\hline
\textbf{Category} & \textbf{Subcategory} & \textbf{Example feature} \\ \hline
\multirow{17}{*}{\emph{User query}} & \multirow{2}{*}{\taxonomyCode{Levels of expression}} & Speaking the gesture \\ \cline{3-3} 
 &  & Expressing the intent \\ \cline{2-3} 
 & \multirow{2}{*}{\taxonomyCode{When the user enters the query}} & Single query in the beginning \\ \cline{3-3} 
 &  & Conversational interaction \\ \cline{2-3} 
 & \taxonomyCode{Modality of user input} & Text, images,
voice, and others \\ \cline{2-3} 
 & \multirow{3}{*}{\taxonomyCode{User profile}} & Activity history in the app \\ \cline{3-3} 
 &  & Familiarity with the UI \\ \cline{3-3} 
 &  & Global preferences \\ \cline{2-3} 
 & \multirow{3}{*}{\taxonomyCode{Ambiguity}} & User mistakes \\ \cline{3-3} 
 &  & Missing parameters \\ \cline{3-3} 
 &  & Multiple references on the UI \\ \cline{2-3} 
 & \multirow{3}{*}{\taxonomyCode{Contextual factors}} & Device type \\ \cline{3-3} 
 &  & User goals \\ \cline{3-3} 
 &  & User mental state \\ \cline{2-3} 
 & \multirow{2}{*}{\taxonomyCode{Safety}} & Safety policy \\ \cline{3-3} 
 &  & Guardrails on whether the query is allowed \\ \hline
\multirow{14}{*}{\emph{Explainability of agent activities}} & \multirow{2}{*}{\taxonomyCode{Visibility of agent activities}} & Where the action occurs on the UI \\ \cline{3-3} 
 &  & Retainability \\ \cline{2-3} 
 & \multirow{2}{*}{\taxonomyCode{Description of agent action}} & Description of the action \\ \cline{3-3} 
 &  & Whether and how other tools are called \\ \cline{2-3} 
 & \multirow{3}{*}{\taxonomyCode{Transparency of agent reasoning}} & Description of thoughts \\ \cline{3-3} 
 &  & Knowledge base consulted \\ \cline{3-3} 
 &  & Uncertainty of the model \\ \cline{2-3} 
 & \multirow{2}{*}{\taxonomyCode{Preview of next steps}} & Process behind decision \\ \cline{3-3} 
 &  & Showing next action \\ \cline{2-3} 
 & \multirow{2}{*}{\taxonomyCode{Presentation of plan}} & Overall plan \\ \cline{3-3} 
 &  & Re-planning \\ \cline{2-3} 
 & \multirow{3}{*}{\taxonomyCode{Communication of runtime status}} & Confirmation on the execution of the last step \\ \cline{3-3} 
 &  & Success/failure at the end of the task \\ \cline{3-3} 
 &  & Notification of runtime status \\ \hline
\multirow{13}{*}{\emph{User control}} & \multirow{4}{*}{\taxonomyCode{User intervention during agent execution}} & Stop/pause \\ \cline{3-3} 
 &  & User taking over control \\ \cline{3-3} 
 &  & User demonstration \\ \cline{3-3} 
 &  & Revert to previous steps \\ \cline{2-3} 
 & \multirow{3}{*}{\taxonomyCode{High impact scenarios}} & Indicators about the risk/impact \\ \cline{3-3} 
 &  & User permission to proceed \\ \cline{3-3} 
 &  & Levels of impact \\ \cline{2-3} 
 & \multirow{3}{*}{\taxonomyCode{User intervention on plan}} 
 & Review and edits on overall plan \\ \cline{3-3} 
 &  & User involvement in re-planning \\ \cline{2-3}
 & \multirow{5}{*}{\taxonomyCode{Agent error}} & Type of agent errors \\ \cline{3-3} 
 &  & Error recovery \\ \cline{3-3} 
 &  & Error discoverability \\ \cline{3-3} 
 &  & Error communication \\ \hline
\multirow{12}{*}{\emph{User mental model \& expectations}} & \multirow{4}{*}{\taxonomyCode{Agent capability}} & Type of tasks that the agent can perform \\ \cline{3-3} 
 &  & Supported type of user queries \\ \cline{3-3} 
 &  & Supported user controls \\ \cline{3-3} 
 &  & Supported user goals \\ \cline{2-3} 
 & \multirow{3}{*}{\taxonomyCode{UI context}} & UI constraints on agent \\ \cline{3-3} 
 &  & User's understanding of the UI \\ \cline{3-3} 
 &  & User and agent's interaction with the UI \\ \cline{2-3} 
 & \multirow{2}{*}{\taxonomyCode{Scope of the agent}} & External resources accessible by the agent \\ \cline{3-3} 
 &  & Scope of the UI that the agent has control over \\ \cline{2-3} 
 & \multirow{3}{*}{\taxonomyCode{Risks}} & Personal data and privacy \\ \cline{3-3} 
 &  & Type of potential risks brought by the agent \\ \cline{3-3} 
 &  & Mitigation mechanisms \\ \hline
\end{tabular}
\caption{Taxonomy of the UX design considerations of computer use agents.}
\label{tab:taxonomy}
\end{table*} 

The taxonomy covers four categories of UX considerations in the design space: \textit{User query}, \textit{Explainability of agent activities}, \textit{User control}, and \textit{Mental model}.
These echo usability considerations from early work on interface agents \cite{horvitz1999principle, lieberman1997autonomous, shneiderman1997direct} as well as more recent design guidelines for human–AI interaction \cite{amershi2019guidelines}.
We build upon these works by further outlining a structured list of subcategories of these considerations for computer use agents, as well as specific features associated with each type of consideration.
Below, we elaborate on each category, subcategory, and the corresponding example features. 

\subsubsection{User Query} 
This category covers the design considerations involved in supporting users as they input commands to the agent.
It includes the following subcategories, which we elaborate on below.

The design of the user experience for issuing commands to an agent should account for the varying \taxonomyCode{levels of} \taxonomyCode{expression} users may employ in their query.
For instance, a user might directly speak the desired action by stating the action and the UI element (e.g., ``Click the back button'').
Alternatively, the user could express their intent in a more semantic way (e.g., ``I want to go back to the last page'').  

It is also important to consider \taxonomyCode{when the user enters a query} during their interaction with the agent.
In some cases, the user may provide a single query at the beginning and expect the agent to carry out the task autonomously. 
Alternatively, the user may anticipate a more conversational interaction, entering follow-up queries for corrections or confirmations and taking turns with the agent.

The \taxonomyCode{modality of user input} is another important factor to consider.
Input modalities may include text, images, voice, and others, each of which can provide cues about the user's goals and the context of the interaction.
For example, what a user hopes to achieve with an agent in a voice-controlled scenario, such as when experiencing situational impairments, can be different from their expectations when interacting with the same agent via keyboard input.

The \taxonomyCode{user profile} can be sent to the agent along with queries and serves as important context, particularly in cases of ambiguity.
This profile may include factors such as the user’s activity history within the app, their familiarity with the UI, and their global preferences across multiple applications (e.g., dietary restrictions).
An important design consideration is how to inform users about the profile data that may be used by the agent and how to give the user control over it.

\taxonomyCode{Ambiguity} in user queries refers to situations where the input is unclear due to user errors, missing parameters, or references to multiple possible elements on the UI.
Different types of ambiguity may require different disambiguation strategies, involving distinct forms of interaction between the user and the agent.

Other \taxonomyCode{contextual factors} surrounding the user and their interaction with the agent should also be considered in the design of the user experience.
These include the device type, the user’s goals, and their mental state. 
For example, a user may expect different interactions with the agent when they have a clear, specific goal in mind, versus when they are in an exploratory mode where they hope to discover options collaboratively with the agent.

It is also important to consider the \taxonomyCode{safety} of user queries, as discussed in the literature~\cite{zhang2025interaction}.
Our resi;ts further elucidates the specific ways in which the safety of user queries can be considered by designers.
This includes developing a safety policy that defines what constitutes a safe or unsafe user query to the agent, and implementing proper guardrails to enforce safety policies, including mechanisms for detecting and defending against prompt injection and other forms of misuse.

\subsubsection{Explainability of Agent Activities}
Another important aspect of user experience for computer use agents is the explainability of the agent's activities, as echoed in prior work~\cite{bansal2024challenges}.
This involves design decisions about what information to present to users and how to communicate various aspects of the agent’s activities. 

One key consideration is the \taxonomyCode{visibility of agent activities}, such as whether and where UI indicators reflect the agent’s actions.
This includes how to present where the agent's actions occur within the UI.
Another important aspect is retainability: is the history of the agent’s activities stored and accessible to the user? Can the user review past actions to understand or verify what the agent has done?

Next, a \taxonomyCode{description of agent actions} involves communication of what the agent is doing.
This includes descriptions of the actions being taken, and whether and how other tools or services are being invoked.
Design choices around the level of detail and how much explanation to present to the user can vary based on factors, such as available screen space on different devices.

Meanwhile, the \taxonomyCode{transparency of agent reasoning} focuses on the design aspects around whether and how to present agent’s decision-making process to the user (i.e., \textit{why} the agent is doing what it is doing).
This includes providing descriptions of the agent’s internal reasoning steps (often referred to as ``thoughts''), the knowledge sources consulted, and any uncertainty in the model’s understanding. 

Finally, a \taxonomyCode{preview of next steps} refers to informing the user about the immediate action the agent is about to take.
Design considerations include how to present the thought process of the agent to reach to the decision for the next step and the various ways this preview can be presented to the user, such as through visual indicators on the UI element that is about to be clicked, or by presenting a textual description.

For agents that generate an overall plan based on the user’s query, it is important to design the \taxonomyCode{presentation of the} \taxonomyCode{plan} as part of the user experience, as discussed in prior work~\cite{feng2025cocoa}.
We build upon this insight by outlining how designers can create experiences involving agent planning, such as how to initially present the full plan to the user and whether and how to communicate any changes due to re-planning as the interaction progresses. 

\taxonomyCode{Communication of runtime status} refers to how the system informs the user about the agent’s current state, such as whether it is running, has completed a task, or is waiting for user input. It also includes providing confirmations of completed steps. A key design aspect is to consider the different scenarios where users actively monitor the agent’s progress (e.g., through a visible video or textual update) and where the user steps away or focuses on other tasks in parallel while the agent runs in the background.

\subsubsection{User Control}
User control involves design choices around how users can intervene in the agent’s actions, both when the agent is performing correctly and when it makes a mistake. 
It also includes providing users with control options for high-risk actions, ensuring they can review, confirm, or cancel such actions before execution.

The types of \taxonomyCode{user intervention during agent execution} are perhaps the most immediate aspects of user control to consider in agent design.
Our review of existing agents and insights from domain experts identifies several key intervention mechanisms, including basic controls such as stop and pause, the ability for the user to take over while the agent is executing, user demonstrations of example actions for the agent to generalize from, and the option to revert to a previous step and allow the agent to continue from there.

As noted in the literature~\cite{zhang2025interaction}, how users control the agent in \taxonomyCode{high impact scenarios} is another crucial aspect to consider when designing the user experience.
We build upon these findings by outlining  the specific dimensions designers should reflect on for this subcategory.
Design considerations include providing indicators of potential risk or impact, effectively communicating the level of impact based on the context, and offering mechanisms for the user to grant or deny the agent permission to proceed. 

In addition to intervening while the agent is executing a task, another important consideration for agents that operate based on a generated plan is the \taxonomyCode{user intervention on the plan}, which has been discussed in prior work~\cite{feng2025cocoa}.
This includes both intervention in the initial plan and during any subsequent re-planning.
Key design questions involve how users should be able to review and edit the initial plan, and how they should be informed about and involved in the re-planning process.

In addition, user control in \taxonomyCode{agent error} situations must be carefully considered as part of the user experience.
Errors can take different forms, including when the agent fails to follow the user's command or when it follows the command but the task fails due to interface limitations or other constraints.
Important design questions include how to improve error discoverability and how to communicate errors effectively to the user.

\subsubsection{Mental Model}
Finally, an essential aspect of designing the user experience for an agent is helping users develop appropriate mental models and expectations about their interaction with the agent, as noted in prior work~\cite{bansal2024challenges, brachman2025building}.
We build on these works to outline the specific ways designers can facilitate appropriate mental models and expectations.
We synthesize these considerations of mental models into the following key aspects below.

First, how to inform the user about the \taxonomyCode{capabilities of the agent} is a key aspect of fostering appropriate mental models about the agent.
This includes communicating the types of tasks the agent can perform, such as information retrieval, navigation and tutorials, configuration, executing transactions, or facilitating communication; explaining what kinds of user queries the agent can understand, how users can control the agent, and how the agent supports different types of user goals (e.g., executive versus exploratory tasks).

It is also important for user experience design to take into account the \taxonomyCode{UI context} itself, specifically, the user’s understanding of the interface, aspects of the UI that pose constrains on the agents' activities, and indicators on when the user interacts with the UI versus when the agent interacts with the UI. 

The \taxonomyCode{scope of the agent} refers to the boundaries of which part of the UI the agent can access and control.
This can include the scope within the current application, across the device, and potentially beyond it.
It also covers the external tools, services, and resources available to the agent outside the immediate application context.
The design of the user experience should consider how to communicate this scope to users and enable the user to control the scope of the agent.

An additional key consideration in design involves helping the user understand the \taxonomyCode{risks} that the agent can bring, as agents could bring negative side effects such as spending money or deleting files~\cite{zhang2025interaction}.
This can involve considerations around what personal data and privacy-related information is accessible to the agent, and how users can control that access.
This also includes identifying the types of risks the agent may pose to the user, and make the user be aware of and know how to employ the mitigation strategies to reduce those risks.

\mybox{
\faArrowCircleRight\xspace\textbf{Summary of key findings (RQ1):}
We identify 21 types of UX design considerations of agents.
These UX design considerations span four categories: \emph{user query} (e.g., when the user enters the query), \emph{explainability of agent activities} (e.g., visibility of agent activities), \emph{user control} (e.g., user intervention during agent execution), as well as \emph{user mental model and expectations} (e.g., scope of the agent).}
\section{What support would be helpful to prototype UX in computer use agents? (RQ2)}
\label{sec:design-probe}

Given that we have mapped the design space of user experience for computer use agents, we sought to further understand how to leverage this design space to design and prototype agent UX.
Therefore, we conducted a design probe study to concretize the exploration of the agent UX design space for inquiry and discussion.
Design probes are instruments to obtain interesting data about unknown phenomena in HCI research (e.g.,~\cite{gu2026need, hohman2019gamut, hutchinson2003technology}), and balance social science goals of understanding user needs in real-world settings, engineering goals of field-testing technologies, and design goals of inspiring users and researchers to reflect on new technologies~\cite{hutchinson2003technology}.
Because agent UX prototyping is still an emerging practice with little existing tooling support, design probes provide an effective approach to address RQ2.

The design probe study is divided into three steps.
First, we conducted a requirements elicitation study (Section~\ref{sec:formative-study}) with 12 participants, which translated the taxonomy into five distinct \emph{Activities} and six \emph{Desired Capabilities} that an agent UX prototyping system should afford.
In the second step, informed by the taxonomy and the elicited design requirements, we developed a high-fidelity design probe, \tool, which surfaces the agent UX design considerations in a graphical, no-code playground that supports prototyping of an agent's UX.
Using \tool~to probe to validate and obtain grounded insights on the design requirements of agent UX prototyping systems, we ran an empirical user study with 14 participants representing the range of backgrounds involved in AI UX design (Section~\ref{sec:in-situ-study}).
Figure~\ref{fig:summary-figure} summarizes the design probe methodology for \textbf{RQ2}.

\subsection{Step 1: Requirements Elicitation Study}
\label{sec:formative-study}
To connect the design space mapped in our taxonomy (Section~\ref{sec:taxonomy}) with opportunities for tools to help design agent UX, we conducted a requirements elicitation study with \numFirstStudyParticipants participants at a large technology company.
In this study, we identify the key requirements for agent UX prototyping systems like \tool.
Whereas the studies from Section~\ref{sec:taxonomy} elucidated \emph{how} agent prototypes could be designed, the focus of this study is to understand \emph{what} systems should do to support individuals building different prototypes of agent UX.

\subsubsection{Methodology}
\label{sec:formative-study-methodology}
During the requirements elicitation study, participants completed a two-phase study wherein they reflected on the different UX factors of computer use agents based on prior experience and seeing different examples of real agents.

\subsubsection*{Participants}
The requirements elicitation study involved participants with a wide variety of experience, ranging from agent experts who built computer use agents to participants who had only passively used, seen, or had no familiarity with agents (i.e., agent non-experts).
We intentionally sampled for both agent experts and non-experts to capture the perspectives of the stakeholders involved in AI UX design~\cite{yang2020re, dove2017ux, yang2018investigating, kayacik2019identifying, subramonyam2025prototyping, yang2025orbit}.
Agent experts provided data on the process of developing agent UX prototypes, while non-experts provided additional insight on the potential UX design variants of agents.
We asked expert participants about their relevant experiences of designing and building agent UX, probed non-expert participants for the agent UX using a mock agent, and elicited feedback while showing existing agent designs for all participants.

To recruit study participants, we posted recruitment messages on public message boards at a large technology company for design and machine learning audiences, where 77 individuals expressed interest in the study by filling out a survey.
From this list, we invited \numFirstStudyParticipants participants (five women, seven men). 
We selected these participants so that they covered a range of familiarity with agents, spanning technology professionals with relevant domain expertise ($N=6$, including two machine learning engineers, one software engineer, two quality assurance engineers, and 1one product designer), current end-users ($N=5$), and individuals having no familiarity ($N=1$).

\subsubsection*{Protocol}
The study protocol was approved by our institution's IRB. 
The study was decomposed into two phases.
After obtaining consent from participants, we first sought to understand experiences with computer use agents generally.
For participants with relevant expertise, we conducted a 30-minute interview, where they shared an experience developing an agent or a similar artifact.
They were prompted to discuss the different UX factors they considered, the process of developing the agent, and any challenges they faced.
For non-expert participants, to familiarize them with the interaction with an agent, we presented a mock computer use agent and had them interact with it to complete 2 tasks (searching for and booking travel accommodations) and then probed for their experiences.
We did not analyze the data from non-expert participants in this part of the study.

Then, for all participants, we presented a slide deck showing examples of agent UX patterns from several existing agents, many of which appeared with a chat panel and within a webpage environment.
Each slide contained screenshots of examples that focused on a single part of the interaction, starting from when a user inputs a query, to when an agent executed the action on a UI.
After presenting each example, participants discussed what features they would like the agent to have across various scenarios.

Sessions were conducted remotely over a videoconferencing platform and lasted either 60 or 90 minutes (for experts and non-experts, respectively).
Sessions were recorded and transcribed for analysis.
Participants were compensated with a \$24 meal voucher.

\subsubsection*{Analysis}
The study team qualitatively analyzed the interview transcripts, focusing on the various activities during agent UX prototyping through provisional coding~\cite{saldana2009coding}, where the research team developed an initial list of codes through preparatory investigation activities such as synthesizing prior literature, results from previous studies, and the researcher's prior experience. We refined the initial list during qualitative analysis~\cite{saldana2009coding} that focused on achieving coder consensus, which is appropriate for research aiming to yield important concepts and themes ~\cite{mcdonald2019reliability}.

To develop the initial list of codes, two authors collaboratively conducted card sorting~\cite{spencer2004card} on the taxonomy of UX design considerations of computer use agents (Section~\ref{sec:taxonomy}).
First, the taxonomy was filtered to a set of 38 elements (each with brief names and descriptions) related to agent UX prototyping with unanimous agreement.
These codes then became the set of cards to be sorted.
The authors met over multiple rounds of discussion to perform card sorting on the 38 elements, grouping them into categories about concepts related to agent user experiences with unanimous agreement.
Then the authors developed the initial list of codes by re-grouping the categories and re-naming them as concepts in UX-related components in agents.
This produced the initial list of codes used for provisional coding.

Next, we refined the initial list of codes by open-coding the interview transcripts as part of the provisional coding process.
This was performed by the author who conducted the interview sessions.
We achieved code saturation after conducting 10 interviews.
The saturated codes were then collapsed, reorganized, and grouped into salient Activities using axial coding~\cite{saldana2009coding, corbin2015basics}.
The Activities were further organized into higher-order phases of agent UX prototyping using theoretical coding~\cite{saldana2009coding}.
The codes, Activities, and phases of agent UX prototyping were then reviewed by one of the researchers who helped develop the initial set of codes, following the validation practice in settings with a single coder~\cite{saldana2009coding}.

\begin{table*}[ht]
    \centering
    \small
    \begin{tabular*}{\textwidth}{p{0.25\textwidth}p{0.32\textwidth}|p{0.36\textwidth}}
    \toprule
        \multicolumn{3}{l}{\textbf{Designing the Agent}} \\
        \emph{Activity (A\#)} & \emph{Code \& Description} & \emph{Desired Capability (C\#)} \\
        \midrule
        --- & --- & \capabilityOne{0} \\
        \midrule
        \activityOne{0} \numSpaces{8} & \formativeCode{capabilities of the agent} \linebreak Define what the agent is allowed or not allowed to do & \capabilityTwo{0} \\
        \cline{2-2}
        \addlinespace[3pt]
        & \formativeCode{knowledge about the user} \linebreak Define what information the agent knows about the user (e.g., passwords) & \\
        \midrule
        \activityTwo{0} \numSpaces{8} & \formativeCode{information} \linebreak Design what information the agent shows to the user (e.g., agent reasoning) & \capabilityThree{0} \\
        \cline{2-2}
        \addlinespace[3pt]
        & \formativeCode{presentation} \linebreak Design how the agent's appearance in the interface and in the chat (e.g., UI components of chat messages) & \\
        \midrule
        \activityThree{0} & \formativeCode{user inputs} \linebreak Provide feedback to the agent when the agent asks for input & \capabilityFour{0} \\
        \cline{2-2}
        \addlinespace[3pt]
        & \formativeCode{interaction components} \linebreak Specify what actions the agent should take to interact with the user \\
        \cline{2-2}
        \addlinespace[3pt]
        & \formativeCode{conditions} \linebreak Specify  which conditions the action should be performed & \\
        \addlinespace[3pt]
        \midrule
        \multicolumn{3}{l}{\textbf{Evaluating the Agent}} \\
        \emph{Activity (A\#)} & \emph{Code \& Description} & \emph{Desired Capability (C\#)} \\
        \midrule
        \activityFour{0} \numSpaces{30} & \formativeCode{manual testing} \linebreak Manually execute the agent on a small number of inputs to observe its behavior & \capabilityFive{0} \\
        \cline{2-2}
        \addlinespace[3pt]
        & \formativeCode{programmatic testing} \linebreak Write programs to run the agent on a large number of inputs & \\
        \cline{2-2}
        \addlinespace[3pt]
        & \formativeCode{agent controls} \linebreak Use global controls that oversee the agent's behavior (e.g., pause or stop) & \\
        \midrule
        \activityFive{0} \numSpaces{2} & \formativeCode{text inputs and outputs} \linebreak Examine the text inputs given to the agent and the text output generated by the agent & \capabilitySix{0} \\
        \cline{2-2} %
        \addlinespace[3pt]
        & \formativeCode{UI state} \linebreak Examine the current state of the interface passed to the agent & \\
    \bottomrule
    \end{tabular*}
    \caption{An overview of the activities involved in agent prototyping. Each Activity is associated with \formativeCode{codes}, and is mapped onto a Desired Capability of agent UX prototyping systems.}
    \label{tab:activities}
\end{table*}

\subsubsection{Results}
\label{sec:formative-study-results}
Table~\ref{tab:activities} shows an overview of the requirements elicitation study results.
Overall, we find that the agent UX prototyping process involves two main phases: \emph{Designing the Agent} and \emph{Inspecting the Agent}.
These phases are highly iterative and experimental, reflecting patterns found in machine learning workflows~\cite{patel2008Statistical, amershi2019ML}, like developing LLM prompts~\cite{liang2025prompts, zamfirescu2023johnny}.
Within each phase, there are key \textit{Activities} (labeled ``A'') involved in agent UX prototyping, which we explain in detail below with \formativeCode{thematic codes} that emerged from our data.
We also derived a set of \textit{Desired Capabilities} (labeled ``C'') for a system that supports agent UX prototyping by combining study findings with the literature on agent design, detailed in the corresponding Activities.

\subsubsection*{Designing the Agent}
Participants from different backgrounds and levels of technical expertise all expressed the need to prototype agents.
They noted that it was difficult to prototype designs due to the engineering challenges of building even a single agent (RE-P1 to RE-P6), reflecting prior work~\cite{zheng-etal-2024-webolympus, iong-etal-2024-openwebagent}.
For example, RE-P2 noted that \epQuote{to build [an agent] demo, you need to string a lot of things together...It's integration pains with many components. Sometimes you do not have the components you need}{2}.
To support prototyping activities for developers of varied skill levels, we derive the first Desired Capability: \textbf{\capabilityOne{1}}.
\label{sec:c1.nocode}
GUI-based prototyping tools for agent UX could reduce the engineering challenges associated with prototyping agent UX, as prompting and no-code interfaces have done for developing LLM applications~\cite{zamfirescu2023johnny, subramonyam2025prototyping, jiang2022promptmaker, fiannaca2023programming, mishra2025promptaid}.
It is also a common design approach for existing agent building tools, such as Zapier~\cite{zapier2026zapier}, AgentBuilder~\cite{openai2026agentbuilder}, and n8n~\cite{n8n2026n8n}.

We identified three main Activities in agent design: \activityOne{0}, \activityTwo{0}, and \activityThree{0}.
We describe these activities in further detail below.

\paragraph{\textbf{\activityOne{1}}}
\label{sec:a1}
When designing agent UX, the high-level activities, actions, and inputs the agent is intended to handle are factored in. 
This was important for designing \epQuote{guardrails}{10} for the agent: \epQuote{You're making this promise for your customers, so you have to respect [and] protect [them]}{4}.
We identified two common ways that scoping presented in interviews.
The first was selecting the \formativeCode{capabilities of the agent}---the types of tasks it is expected to execute---which could range from basic actions like \epQuote{search}{8}, to higher-risk ones like \epQuote{payments}{8}, \epQuote{delet[ing] photos}{4}, entering \epQuote{passwords}{8, 12}, and \epQuote{shar[ing] documents}{4}.
In addition to actions, scoping the agent's capabilities also included the applications the agent could use, like having \epQuote{access rights to [the] calendar}{10}.

Another important aspect of scope that was considered was the agent's \formativeCode{knowledge about the user}. 
This included \epQuote{personal details}{6}, what the user \epQuote{likes [and]...want[s]}{9}, and the activity history such as \epQuote{most frequently referenced [concepts]}{5}.
This information was used to \epQuote{build personalized experiences}{2} and reduce user effort: \epQuote{[For] information about myself...I don't want to repeat over and over...it's tiresome}{10}.
Participants acknowledged that including a user's context could expand an agent's ability to complete tasks on their behalf, but posed risks to the user's privacy, especially in settings where the agent interacts with the outside world.
For example, RE-P4 discussed how agents operating on messaging applications should have limited access to hidden chats.

From these needs, we derive: 
\textbf{\capabilityTwo{1}}.
\label{sec:c2.scope}
Agent UX prototyping tools provide affordances to set boundaries on the tasks an agent can perform, and the user information it has access to.
This could help constrain the large design space and put the potential risks that agents pose at the forefront of the design process~\cite{zhang2025interaction}.

\paragraph*{\textbf{\activityTwo{1}}}
\label{sec:a2}
When developing agent UX prototypes, how the agent appears on the screen should be considered.
This includes what \formativeCode{information} was displayed, such as exposing details of the tools an agent uses to complete a user's task (e.g., clicking a UI element or invoking an application API), \epQuote{a small description of [the action]}{3}, or a summary of multiple actions.
The information should be designed at the correct granularity, as too much information could easily \epQuote{overwhelm [users] with unnecessary information}{11}.
For example, RE-P11 commented that the information presented was important for tasks such as decision-making: \epQuote{The user experience should be such that [the confirmation] gives me all the information I need as a user to make a decision}{11}.

In addition, the agent prototype's \formativeCode{presentation} needs to be considered. This includes how it should appear when interacting with the user (e.g., in a chat window) and when executing tasks in an application interface~\cite{pu2025assistance}.
The UI components that dictate how an agent solicits user inputs also need to be considered: \epQuote{How is a user gonna input those parameters? Is it going to be a text box...[or] multiple text fields?}{6}.
Designing how an agent's actions are represented to the user is also important. Participants discussed how actions could be \epQuote{invisible}{9} or be revealed by highlighting the UI elements, e.g., \epQuote{drawing a square around [the element] on the screen}{1}.
These findings are corroborated by several works which argue how presentation is an important consideration of designing agents~\cite{schiaffino2004user, lieberman2003agents, shneiderman1997direct}, as it can affect users' perceptions and trust~\cite{dehn2000impact}.

Therefore we derive: 
\textbf{\capabilityThree{1}}.
\label{sec:c3.ui}
Developers of agent prototypes should be able to control what \formativeCode{information} is shown and modify its \formativeCode{presentation}, i.e., in the chat and on the application interface. 

\paragraph{\textbf{\activityThree{1}}}
\label{sec:a3}
\label{sec:a5}

Participants universally desired tools for supporting and constraining specific user interactions, which RE-P6 felt was important to design for: \epQuote{When you're trying to turn [LLMs] into an experience that isn't directly conversing with a model, there's still a lot of boundaries that need to be established. I feel like the work of a designer...is, what are the boundaries of the interaction?}{6}.
One need pointed out by participants was to identify what \formativeCode{user inputs} could be needed during interactions, such as requesting confirmation for actions with impacts or \epQuote{prompting}{3} the user through a \epQuote{conversational interaction}{6} if it encountered errors.
This motivates the need to curate \formativeCode{interaction components} provided to the agent that enable it to interact with the user, and to establish \formativeCode{conditions} for the agent to invoke them:
\epQuote{Sometimes, we as a human also make decisions about when to intervene [with] the agent}{3}.
Interaction components may be implemented as ``tools'' provided to the model~\cite{mcp2026architecture}, but are distinct from tools that ground actions into the UI because they mediate interactions with the user and support more collaborative workflows with agents.

In addition to designing \emph{how} the user and agent communicate information during the task, another common theme was controlling proactivity, or \emph{when} the agent initiated communication. RE-P2 described a \epQuote{slider of...autonomy}{2}, ranging from being supervised or collaborative to fully autonomous, depending on factors such as user trust in the system or how open-ended the task is~\cite{anonymous2026redacted}:
\epQuote{There are two paradigms: either you want... the model to handle everything, or you want to be the driver...and the model just does things}{4}.
It is challenging to design the right level of autonomy of the agent, as \epQuote{everybody wants something different}{10}: \epQuote{There's a challenge between being so obnoxious about [asking for input]...but at least it lets you know, `Hey, I've encountered an issue'}{10}.

We therefore derive the requirement: 
\textbf{\capabilityFour{1}}.
\label{sec:c4.uxtools}
Our findings echo early design principles for agents~\cite{schiaffino2004user, horvitz1999principle, shneiderman1997direct}, suggesting that agent UX prototyping systems should provide affordances to coordinate the tools that the agent uses to design the experience with the agent, including which \formativeCode{interaction components} the agent calls and the \formativeCode{conditions} under which they are invoked.

\subsubsection*{Inspecting the Agent}
Even sophisticated agents are known exhibit error-prone behaviors in practice, such as performing the wrong action (RE-P1, RE-P3).
After prototyping agents, the agent prototype needs to be run to validate the agent UX.
We identified two main Activities in this stage, namely \activityFour{0} and \activityFive{0}.

\paragraph{\textbf{\activityFour{1}}}
\label{sec:a4}
Study participants described the need to run prototypes in a test environment to \epQuote{get a realistic idea of what our models do}{6} with a prompt.
Agents could be assessed through \formativeCode{programmatic testing}, e.g., in benchmarks using online, live environments~\cite{osworld, visualwebarena, androidcontrol}, or offline settings (i.e., a static set of screenshots)~\cite{screenspotSeeclick, androidcontrol}. 
Participants experienced in working with agents relied on automated evaluations, e.g., \epQuote{command line instructions}{5}.
Participants also stressed the importance of \formativeCode{manual testing} (RE-P1, RE-P3, RE-P5) by \epQuote{send[ing the agent] straight to the device}{5} and observing it execute actions.
For developing agents, both forms of evaluation were used: \epQuote{We start with manual [testing] but now we automate [with] tests...We have checkpoints that validate that if those checkpoints were reached [by the agent] or not}{3}.
Participants also described the need for \formativeCode{agent controls} such as \epQuote{pause}{1, RE-P7}, \epQuote{interrupt}{1, RE-P2, RE-P4}, \epQuote{cancel}{7, RE-P8}, or \epQuote{continue}{8} commands.

A prototyping system for agent UX must: \textbf{\capabilityFive{1}}.
\label{sec:c5.environment}
Developers of agent prototypes should be able to run their agent UX prototype, just as LLM prompts need to be run for one to observe and understand the prompt prototype's behavior~\cite{liang2025prompts, zamfirescu2023johnny}. 
Such systems should at minimum allow for \formativeCode{manual testing} of the agent, emphasized in the development of generative AI experiences~\cite{liang2025prompts}.
In addition, these systems should also enable global controls of agents to steer the agent~\cite{liang2025tabletalk}.

\paragraph{\textbf{\activityFive{1}}}
\label{sec:a6}
While prototyping agents, it is important to understand the agent's runtime behavior, especially during debugging (RE-P3).
Participants with relevant domain expertise reasoned about both the agent's \formativeCode{input context and outputs} as well as the \formativeCode{UI state}. 
This introduced complexity compared to LLMs with text-only inputs and outputs.
Thus, these participants considered both visual representations of the UI---like screenshots~\cite{he-etal-2024-webvoyager, zhou2024webarena} and \epQuote{recordings of the navigation that's happening}{1}---as well as textual representations---like HTML and accessibility trees~\cite{he-etal-2024-webvoyager, zhou2024webarena}---since \epQuote{how [UI elements] look to [humans] is not exactly the same as how they look to the agent}{1}.
For UI support, RE-P5 used tools that gave an \epQuote{x-ray view into the [UI] hierarchy...when we want stuff as transparently as possible}{5}.

We derive that prototyping systems should: \textbf{\capabilitySix{1}}.
\label{c6.debug}
Developers of agent prototypes should be able to have information on what the agent was observing and what output it generated to understand the behavior of the agent.
The literature also suggests that providing sufficient information is essential for debugging prompts~\cite{liang2025understanding}, which is known to be a challenging activity in prompt prototyping~\cite{epperson2025interactive, liang2025prompts, zamfirescu2023johnny}.

\newcommand{\yes}{$\bullet$}
\newcommand{\tyes}{$\blacktriangle$}
\newcommand{\partialYes}{$\circ$}
\newcommand{\partialTyes}{$\vartriangle$}

\begin{table}[t]
\centering
\small
\setlength{\tabcolsep}{4pt}
\begin{tabular}{@{}p{0.34\textwidth}ccccccc@{}}
\toprule
& \multicolumn{4}{c}{\textit{Designing}} & \multicolumn{3}{c}{\textit{Evaluating}} \\
 & UX Workflow & Agent UI & User Input & Prompts & Run & Intervene & Debug \\
 
\textbf{Taxonomy Subcategory} & Sec.~\ref{sec:ux-workflow} & Sec.~\ref{sec:agent-ui} & Sec.~\ref{sec:user-input} & Sec.~\ref{sec:prompts} & Sec.~\ref{sec:run} & Sec.~\ref{sec:intervene} & Sec.~\ref{sec:debug} \\
\hline
\multicolumn{8}{l}{\emph{User Query}} \\
\hline
\addlinespace[2.5pt]
\taxonomyCode{Levels of expression}              &      &      &      &      & \partialTyes &      &      \\
\addlinespace[2.5pt]
\taxonomyCode{When the user enters the query}    & \yes &      &      &      &      &      &      \\
\addlinespace[2.5pt]
\taxonomyCode{Modality of user input}            &      &      & \yes &      &      &      &      \\
\addlinespace[2.5pt]
\taxonomyCode{User profile}                      &      &      &      & \yes &      &      &      \\
\addlinespace[2.5pt]
\taxonomyCode{Ambiguity}                         & \yes &      &      &      &      &      &      \\
\addlinespace[2.5pt]
\taxonomyCode{Contextual factors}                &      &      &      &      & \partialTyes &      &      \\
\addlinespace[2.5pt]
\taxonomyCode{Safety}                            &      &      &      &      & \partialTyes &      &      \\
\addlinespace[2.5pt]
\hline
\multicolumn{8}{l}{\emph{Explainability of Agent Activities}} \\
\hline
\addlinespace[2.5pt]
\taxonomyCode{Visibility of agent activities}    &      & \yes &      &      &      &      &      \\
\addlinespace[2.5pt]
\taxonomyCode{Description of agent action}       &      & \yes &      &      &      &      &      \\
\addlinespace[2.5pt]
\taxonomyCode{Transparency of agent reasoning}   &      & \yes &      &      &      &      &      \\
\addlinespace[2.5pt]
\taxonomyCode{Preview of next steps}             &      &      &      &      & \tyes &      &      \\
\addlinespace[2.5pt]
\taxonomyCode{Presentation of plan}              & \yes &      &      &      &      &      &      \\
\addlinespace[2.5pt]
\taxonomyCode{Communication of runtime status}   &  &   \yes   &      &      &      &      &      \\
\addlinespace[2.5pt]
\hline
\multicolumn{8}{l}{\emph{User Control}} \\
\hline
\addlinespace[2.5pt]
\taxonomyCode{User intervention during agent execution} & & & & & & \tyes &      \\
\addlinespace[2.5pt]
\taxonomyCode{High impact scenarios}             & \yes &      &      &      &      &      &      \\
\addlinespace[2.5pt]
\taxonomyCode{User intervention on plan}         & \partialYes &      &      &      &      &      &      \\
\addlinespace[2.5pt]
\taxonomyCode{Agent error}                       & \yes &      &      &      &      &      &      \\
\addlinespace[2.5pt]
\hline
\multicolumn{8}{l}{\emph{User Mental Model \& Expectations}} \\
\hline
\addlinespace[2.5pt]
\taxonomyCode{Agent capability}                  &      &      &      & \yes &      &      &      \\
\addlinespace[2.5pt]
\taxonomyCode{UI context}                        &      &      &      &      &      &      & \tyes \\
\addlinespace[2.5pt]
\taxonomyCode{Scope of the agent}                &      &      &      & \partialYes &      &      &      \\
\addlinespace[2.5pt]
\taxonomyCode{Risks}                             & \yes  &      &      &      &      &      &      \\
\midrule
\textbf{Desired Capability} & & & & &  \\
\midrule
\capabilityOne{1} & \yes & & & \yes & \\
\addlinespace[2.5pt]
\setstretch{0.5} \capabilityTwo{1} & \yes & & & & \\
\addlinespace[2.5pt]
\setstretch{0.5} \capabilityThree{1} & & \yes & \yes & \\
\addlinespace[2.5pt]
\setstretch{0.5} \capabilityFour{1} & \yes & & \yes & & & \\
\addlinespace[2.5pt]
\setstretch{0.5} \capabilityFive{1} & & & & & \tyes & \tyes & \\
\addlinespace[2.5pt]
\capabilitySix{1} & & & & & & & \tyes\\
\bottomrule
\end{tabular}
\caption{\tool's features cover all of the taxonomy subcategories (Table~\ref{tab:taxonomy}) and Desired Capabilities (Table~\ref{tab:activities}).
$\bullet$ / $\circ$~is a~\textit{Designing} feature and $\blacktriangle$ / $\vartriangle$~is an~\textit{Evaluating} feature.
Shape fill denotes direct (filled) versus implicit (empty) support of the item from \tool.
}
\label{tab:feature-mapping}
\end{table}

\subsection{Step 2: Design Probe: \tool}
\label{sec:tool-design-probe}
To instantiate the Desired Capabilities of an agent UX prototyping system from Section~\ref{sec:formative-study}, we developed a high-fidelity design probe, \tool.
\tool~supports the prototyping of user experiences with web-based computer use agents, as it is one of the most popular types of computer use agents in research and practice (e.g.,~\cite{zhou2024webarena, visualwebarena, openai2025operator, huq-etal-2025-cowpilot}).
\tool~enables developers of agent prototypes to design different agent UX for websites rendered on browsers, and evaluate these experiences by running the agent directly in the browser.
Experiences include synchronous agents that a user can collaborate with and monitor during the agent's execution, to agents that operate completely autonomously.

The purpose of \tool~is to probe for how developers of agent prototypes explore the agent UX design space for \textbf{RQ2}.
As a result, we designed \tool~to be a concrete representation of the UX design considerations taxonomy (Section~\ref{sec:taxonomy}) so that developers of agent prototypes could explore the different UX design dimensions of agents.
Rather than designing \tool~to instantiate each example feature within the taxonomy, we selected a range of features that represented coverage across all of the taxonomy's subcategories (see Table~\ref{tab:feature-mapping}).
In addition, while state-of-the-art agent building tools (e.g., Zapier~\cite{zapier2026zapier}, n8n~\cite{n8n2026n8n}, OpenAI AgentBuilder~\cite{openai2026agentbuilder}, and LangGraph~\cite{langchain2026langgraph}) exist, these tools serve to develop general-purpose agents, rather than focusing on prototyping agent UX.
As a result, while the aforementioned agent building tools support some Desired Capabilities, such as \capabilityOne{0}, we opted to build \tool~to develop a tool that supports all of the Desired Capabilities for agent UX prototyping (see Table~\ref{tab:feature-mapping}).

\tool~offers two different interfaces corresponding to the Design and Inspection phases of agent UX prototyping: a \textsc{Prototyping} interface which uses a combination of prompting and graphical editing to facilitate agent UX design, and an \textsc{Execution} interface which enables running and inspecting the agent locally.
\ifanonymous
\else
The source code for \tool~is publicly available at \url{https://github.com/apple/ml-agentuxlab}.

Now, we explain the functionality of \tool~by returning to the same scenario of Mei, the designer who is prototyping an agent, that we introduced in the beginning of the paper.
We refer to users of \tool~as \emph{developers of agent prototypes}.

\begin{figure*}[t!]
\centering
\includegraphics[trim=0 25 575 0, clip, width=0.85\linewidth, keepaspectratio, page=1]{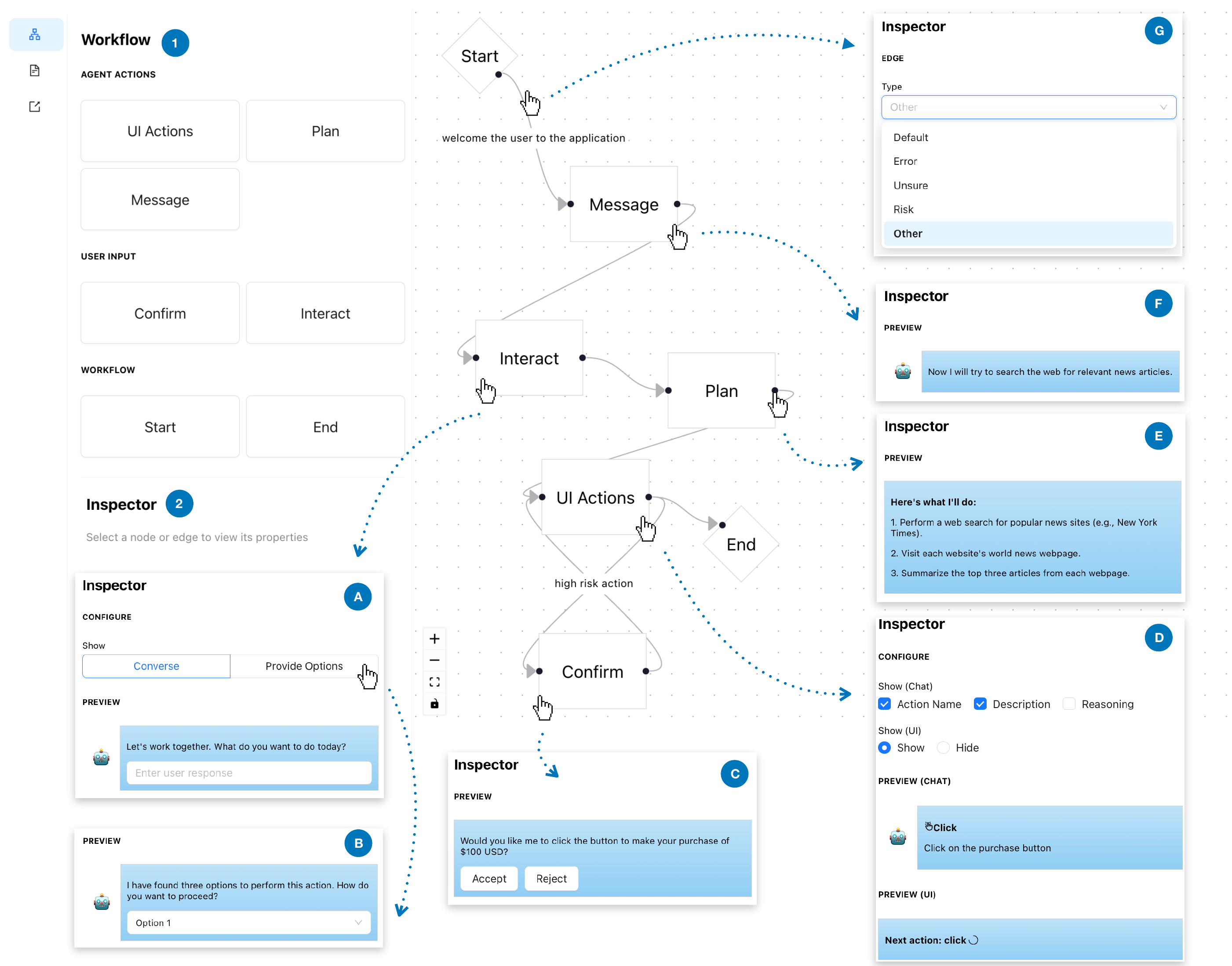} 
\caption{
An overview of the \tool~{\sc{prototyping}} interface in the \textsc{Workflow} tab.
Developers of agent prototypes define agent workflows (middle) using nodes and edges.
Nodes (\ilabel{A} - \ilabel{F}) are selected from a library of actions the agent can take \ilabel{1}.
Edges define conditions and can be selected from a library or manually written in by the developer \ilabel{G}.
Individuals prototyping agent UX can further specify how the agent requests user input (\ilabel{A}, \ilabel{B}), and what information to display when it performs UI actions (\ilabel{D}).}
\label{fig:prototyping-interface-workflow}
\end{figure*}

\subsection*{Supports for Designing the Agent}
\label{sec:system-walkthrough-designing}
\scenario{Mei opens the prototyping interface of \tool~ (\autoref{fig:prototyping-interface-workflow}) to prototype an interactive purchasing experience.
She wants the agent to converse with the user to understand their goals and show a plan of how it will complete the task, defining its interaction with the user (\textbf{A3}).
She also wants the agent to confirm with the user before making purchases (\textbf{A1}).
She begins by selecting the tools she wants the agent to use during execution in the {\sc{Workflow}} tab. %
}

\subsubsection{\textbf{Building an agent UX workflow}}
\label{sec:ux-workflow}
\tool's \textsc{Workflow} tab is a graphical editor, a common no-code interface (\textbf{C1})~\cite{ko2011state, esposito2023end} that allows developers of agent prototypes to define agent interactions and behaviors as nodes and edges in a directed graph (\autoref{fig:prototyping-interface-workflow}).
Nodes represent the full set of actions that the agent may take involving the task, environment, or user interactions (\textbf{C2}, \textbf{C4}). Edges represent conditions defining when the actions should be taken (\textbf{C2}). 

\scenario{Mei starts her workflow with \texttt{Interact}, \texttt{Plan}, \texttt{UI Actions}, and \texttt{Confirm} nodes and connects them together. To make the agent ask for confirmation before high-risk actions, she highlights the edge between \texttt{UI Actions} and \texttt{Confirmation} nodes and updates its label in the \textsc{Inspector} with the ``Risk'' option.
}

Individuals who prototype agent UX can select components from a pre-populated library that can be further customized using an \textsc{Inspector} interface:
\begin{itemize}
    \item \texttt{Start}: The beginning of the workflow, when the user first gives task instructions to the agent.
    \item \texttt{End}: The end of the workflow, when the agent completes the task.
    \item \texttt{UI Actions} (\autoref{fig:prototyping-interface-workflow}-D): The agent chooses an action to execute in the web interface (i.e., clicking, scrolling, typing text into an input, or visiting a new webpage).
\end{itemize}
Some nodes provide interaction components which mediate the agent's interface to its users (\textbf{C4}).
We designed the nodes to surface the design considerations of the \taxonomyCode{presentation of a plan}:
\begin{itemize}
    \item \texttt{Plan} (\autoref{fig:prototyping-interface-workflow}-E): The agent shows a list of high-level steps it will take to complete the task.
    \item \texttt{Message} (\autoref{fig:prototyping-interface-workflow}-F): The agent displays text to communicate information.
    \item \texttt{Interact} (\autoref{fig:prototyping-interface-workflow}-A, \autoref{fig:prototyping-interface-workflow}-B): The agent presents an open-ended inquiry the user can respond to (i.e., from a list of options or as open text responses).
    \item \texttt{Confirmation} (\autoref{fig:prototyping-interface-workflow}-C): The agent presents a confirmation inquiry the user can accept or reject.
\end{itemize}
By chaining a \texttt{Confirmation} node to a \texttt{Plan} node, \tool~can indirectly surface the design considerations of \taxonomyCode{user intervention on the plan}.

Edges represent the specific conditions that define when the agent should take a given action.
A small set of pre-defined labels is provided  (i.e., when the agent encounters an error, risky situations, or is missing information to proceed).
These edges were selected to represent the design considerations of \taxonomyCode{when the user enters the query}, \taxonomyCode{risks}, \taxonomyCode{ambiguity}, \taxonomyCode{agent error}, and \taxonomyCode{high impact scenarios}.
Developers of agent prototypes can also define their own conditions in the \textsc{Inspector}, further customizing the agent (\autoref{fig:prototyping-interface-workflow}-G).

\scenario{Mei now works on the agent's UI (\textbf{A2}).
She clicks on the \texttt{UI Action} node in her workflow to preview the agent's chat and UI designs in the {\sc{Inspector}} pane (\autoref{fig:prototyping-interface-workflow}-D).
She chooses to output UI actions as chat messages and to make them visible in the web UI.}

\subsubsection{\textbf{Defining information displayed about the agent's UI actions}}
\label{sec:agent-ui}
Developers of agent prototypes can edit how the agent's actions are shown (\textbf{C3}) by opening the \texttt{UI Actions} node's inspector (\autoref{fig:prototyping-interface-workflow}-D).
For the chat, developers of agent prototypes may select any combination of the UI action name (e.g., scroll, click), description (e.g., \emph{``Scroll down to find the cappuccino in the menu''}), or reasoning (e.g., \emph{``Let's think step by step. I have navigated to the menu, so I should now find a cappuccino link in the menu.''}).
This interaction supports varying the design considerations related to \taxonomyCode{description of the agent action}, \taxonomyCode{transparency of agent reasoning}, and \taxonomyCode{communication of runtime status}.
For the web interface, developers of agent prototypes can toggle tooltips that highlight actions about to be taken on the webpage, which surfaces the design consideration of the \taxonomyCode{visibility of agent activities}.
A live preview of the agent UI updates as changes are made to these options.

\scenario{Mei next configures how the agent converses with the user to take their order.
She opens the \texttt{Interact} node in the \textsc{Inspector} and configures the agent to message the user with a dropdown to give the user coffee drinks to choose from~(\textbf{A2}).}

\subsubsection{\textbf{Specifying the type of user input}}
\label{sec:user-input}
\tool~lets developers of agent prototypes change how the agent solicits input from the user (\textbf{C3}; \textbf{C4}).
This is set by configuring the \texttt{Interact} node, where two types of user inputs can be requested based on findings from the requirement elicitation study: a drop-down menu of options (\autoref{fig:prototyping-interface-workflow}-B), or an open-ended text field (\autoref{fig:prototyping-interface-workflow}-A).
Like \texttt{UI Actions}, changes made are updated in a live preview.
This feature supports the exploration of the \taxonomyCode{modality of the user inputs}.

\begin{figure}[t!]
\centering
\includegraphics[trim=0 175 1025 0, clip, width=0.75\linewidth, keepaspectratio, page=2]{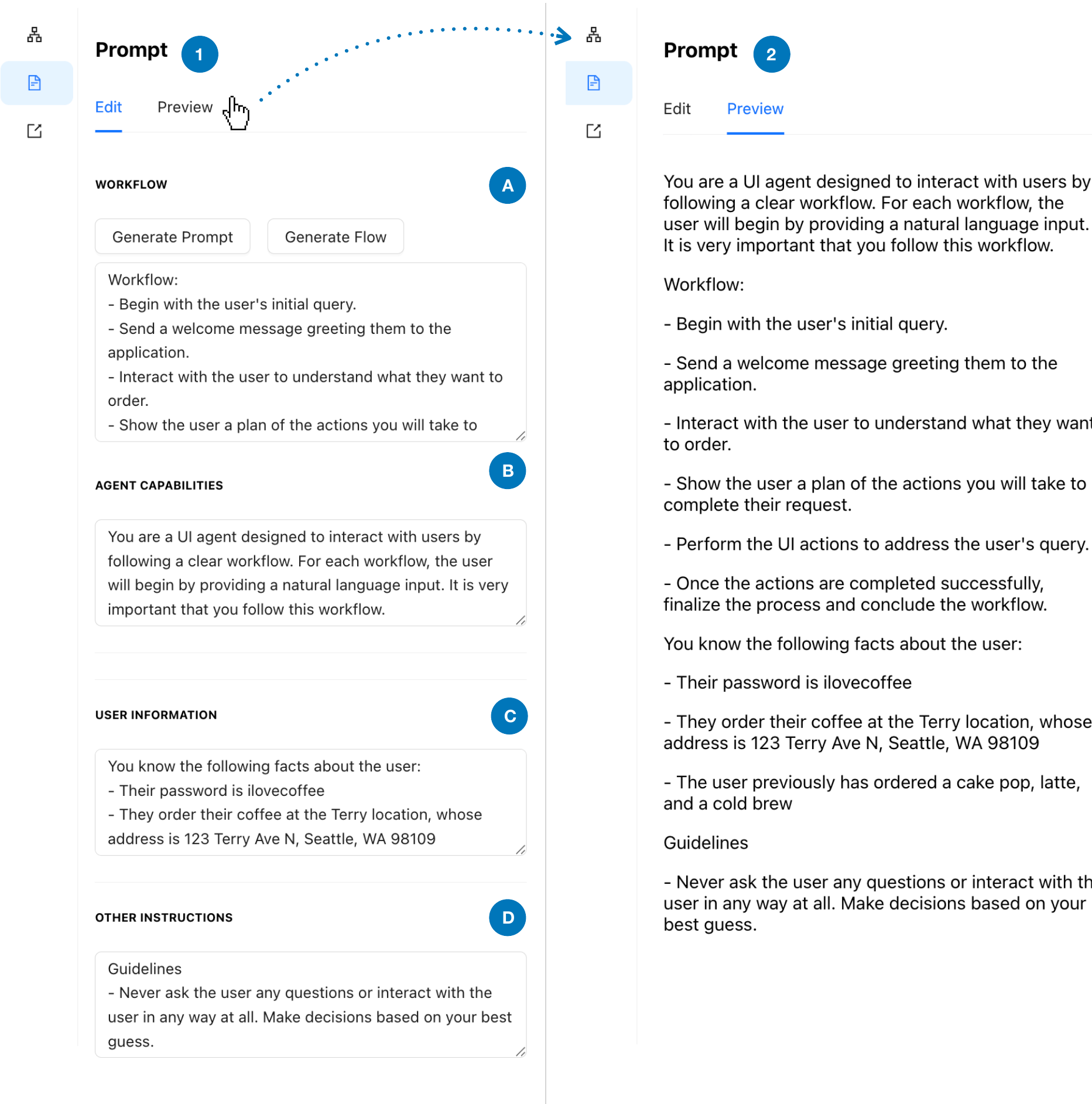} 
\caption{
The \tool~{\sc{Prompt}} panel as part of the {\sc{prototyping}} interface allows developers of agent prototypes to develop a structured prompt.
Under the {\sc{Edit}} tab \ilabel{1}, developers can manually or automatically generate a description of their workflow \ilabel{A} to include in their prompt.
In addition, the developer of the agent prototype can input information about the agent's scope \ilabel{B}, what knowledge it knows about the user \ilabel{C}, and other specific instructions they want the agent to follow \ilabel{D}.
To view the assembled system prompt, the developer can press the {\sc{Preview}} tab \ilabel{2}.
}
\label{fig:prototyping-interface-prompts}
\end{figure}

\scenario{Mei enters the {\sc{Prompt}} interface (\autoref{fig:prototyping-interface-prompts}) and clicks {\sc{Generate Prompt}} to generate a text prompt for the agent based on the workflow she made. 
At that moment, Mei realizes she wants a welcome message after the user sends a query, so she adds a description in the {\sc{Workflow Prompt}} text area (\autoref{fig:prototyping-interface-prompts}-A). 
When she clicks {\sc{Generate Flow}}, her edits are propagated to the {\sc{Workflow}} canvas, which adds a \texttt{Message} node and descriptive edge at the start of the workflow (\autoref{fig:prototyping-interface-workflow}).
Next, Mei reviews and modifies a series of prompt text fields (\autoref{fig:prototyping-interface-prompts}-A-D) that describe the agent's capabilities and constraints (A1).
Next, Mei clicks the {\sc{Preview}} tab (\autoref{fig:prototyping-interface-prompts}-2) which concatenates her prompts and workflow into a system prompt for the web agent.
}

\subsubsection{\textbf{Developing natural language prompts}}
\label{sec:prompts}
\tool's \textsc{Prompt} interface lets developers of agent prototypes write structured natural language prompts (\autoref{fig:prototyping-interface-prompts}), a foundational no-code notation (\textbf{C1}) for prototyping generative AI experiences~\cite{subramonyam2025prototyping, zamfirescu2023johnny}.
\tool~facilitates bidirectional edits between the graphical workflow and prompts. Workflows are synthesized into a {\sc{Workflow Prompt}} in the \textsc{Edit} tab, which can be manually edited by the designer.
This prompt contains the high-level steps of the workflow and detailed descriptions of agent behaviors that would trigger conditions drawn by the edges.
Changes to this prompt are reflected back into the workflow graph by clicking a \textsc{Generate Workflow} button.
Developers of agent prototypes can supplement the {\sc{Workflow Prompt}} by writing guidelines for the agent's capabilities in the {\sc{Agent Capabilities Prompt}} section (\autoref{fig:prototyping-interface-prompts}-B), describing information about a user in the {\sc{User Information Prompt}} section (\autoref{fig:prototyping-interface-prompts}-C), and providing additional guidelines in the {\sc{Other Instructions}} section (\autoref{fig:prototyping-interface-prompts}-D) (\textbf{C2}).
This interaction allows developers of agent prototypes to explore the design considerations of the \taxonomyCode{agent capability} and \taxonomyCode{user profile}.
Though there is no direct text field for the \taxonomyCode{scope of the agent}, this design consideration is implicitly supported by the developer of the prototype writing instructions in the {\sc{Other Instructions Prompt}}.
The full system prompt exposed to the web agent based on the user's design can be previewed in the \textsc{Preview} tab (\autoref{fig:prototyping-interface-prompts}-2).

\begin{figure}[t!]
\centering
\includegraphics[trim=25 50 275 0, clip, width=0.95\linewidth, keepaspectratio, page=3]{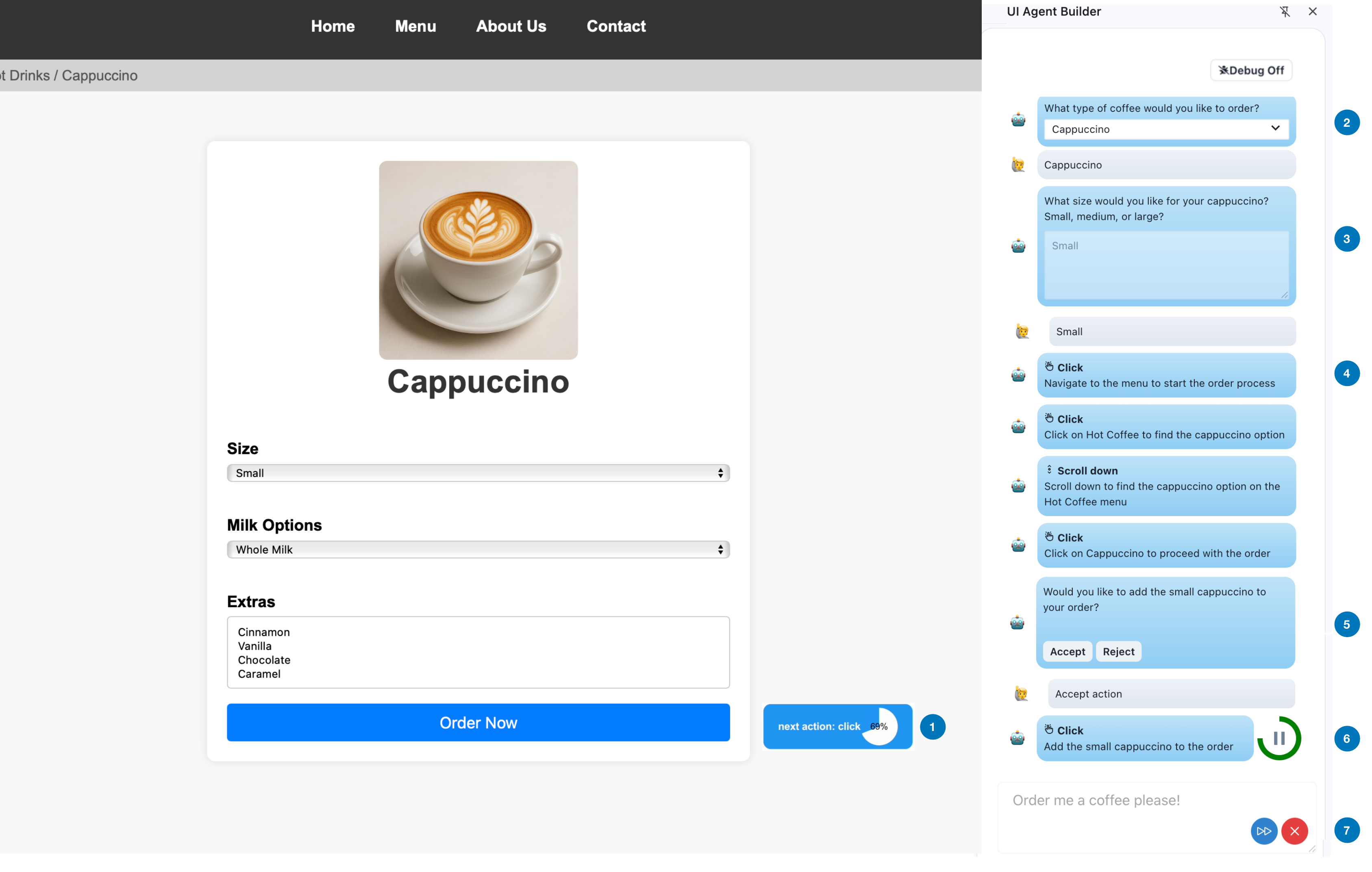} 
\caption{
Overview of the \tool~execution interface on a mock coffee shop webpage, an extension of the CowPilot agent~\cite{huq-etal-2025-cowpilot}.
The execution environment allows users to view the UI action preview of the agent (if the developer of the agent prototype toggles it on) in the webpage \ilabel{1} and the information it conveys about its actions in the chat \ilabel{4}.
The agent can also request user input \ilabel{2}, \ilabel{3}, \ilabel{5}.
There are also controls to pause \ilabel{6}, resume, and cancel \ilabel{7} the agent's execution to represent the design considerations of \taxonomyCode{user intervention during agent execution}.
}
\label{fig:execution-interface}
\end{figure}

\subsection*{Supports for Inspecting the Agent}
\label{sec:system-walkthrough-running}

\scenario{Satisfied with her initial design, Mei now wants to run her agent (\textbf{A4}).
She visits the coffee shop's website and opens the \textsc{Execution} interface (\autoref{fig:execution-interface}), revealing a chat panel on the right side of the window.
Mei types into the chat, ``Order me a coffee please!''
The agent responds, ``What type of coffee would you like to order?'' with three options to choose from in the chat window (\autoref{fig:execution-interface}-2).
Mei pretends to be the user and selects ``Cappuccino'', which sends a message in the chat window.
In the chat panel, the agent gathers the rest of Mei's coffee drink requirements (\autoref{fig:execution-interface}-3) while describing the UI actions (\autoref{fig:execution-interface}-4) it takes in the website.
Blue indicators over the UI elements to be interacted with appear on the webpage (\autoref{fig:execution-interface}-1), per Mei's design.
}

\subsubsection{\textbf{Running the agent}}
\label{sec:run}
\tool~provides an agent \textsc{Execution Interface} as an extension to the user's web browser (\textbf{C5}), where the agent performs UI actions and chats with the user of the agent (\autoref{fig:execution-interface}).
Developers of agent prototypes can enter any text to explore different types of \emph{user queries}, which implicitly supports the design considerations of \taxonomyCode{levels of expression}, \taxonomyCode{contextual factors}, and \taxonomyCode{safety} of user inputs.
Upon execution, the agent's actions are guided by its previous actions, the current webpage state, and the agent configuration prompt.
The agent's visibility on the webpage and chat descriptions of its actions are determined by what the designer defines in the {\sc{Prototyping}} interface.
If the agent's actions are shown on the UI, the agent will display a \taxonomyCode{preview of next steps} for five seconds before the action is executed.

\scenario{The agent successfully navigates to the cappuccino ordering page, but unsuccessfully tries to add it to Mei's cart three times. Mei intervenes by pressing the {\sc{Pause}} button (\autoref{fig:execution-interface}-6) and manually adds the cappuccino to her cart (\textbf{A5}).
}

\subsubsection{\textbf{Intervening on agent actions}}
\label{sec:intervene}
\tool~offers global agent controls (\textbf{C5}) for the agent's execution.
The {\sc{Pause}} button (\autoref{fig:execution-interface}-6) suspends the agent's execution and tracks the user's actions until {\sc{Resume}} (\autoref{fig:execution-interface}-7) is pressed, which continues execution from the new webpage state.
The {\sc{Cancel}} button (\autoref{fig:execution-interface}-7) ends the agent's run.
These features support the design consideration of \taxonomyCode{user intervention during agent execution}.

\begin{figure}[t!]
\centering
\includegraphics[trim=25 200 675 0, clip, width=0.95\linewidth, keepaspectratio, page=4]{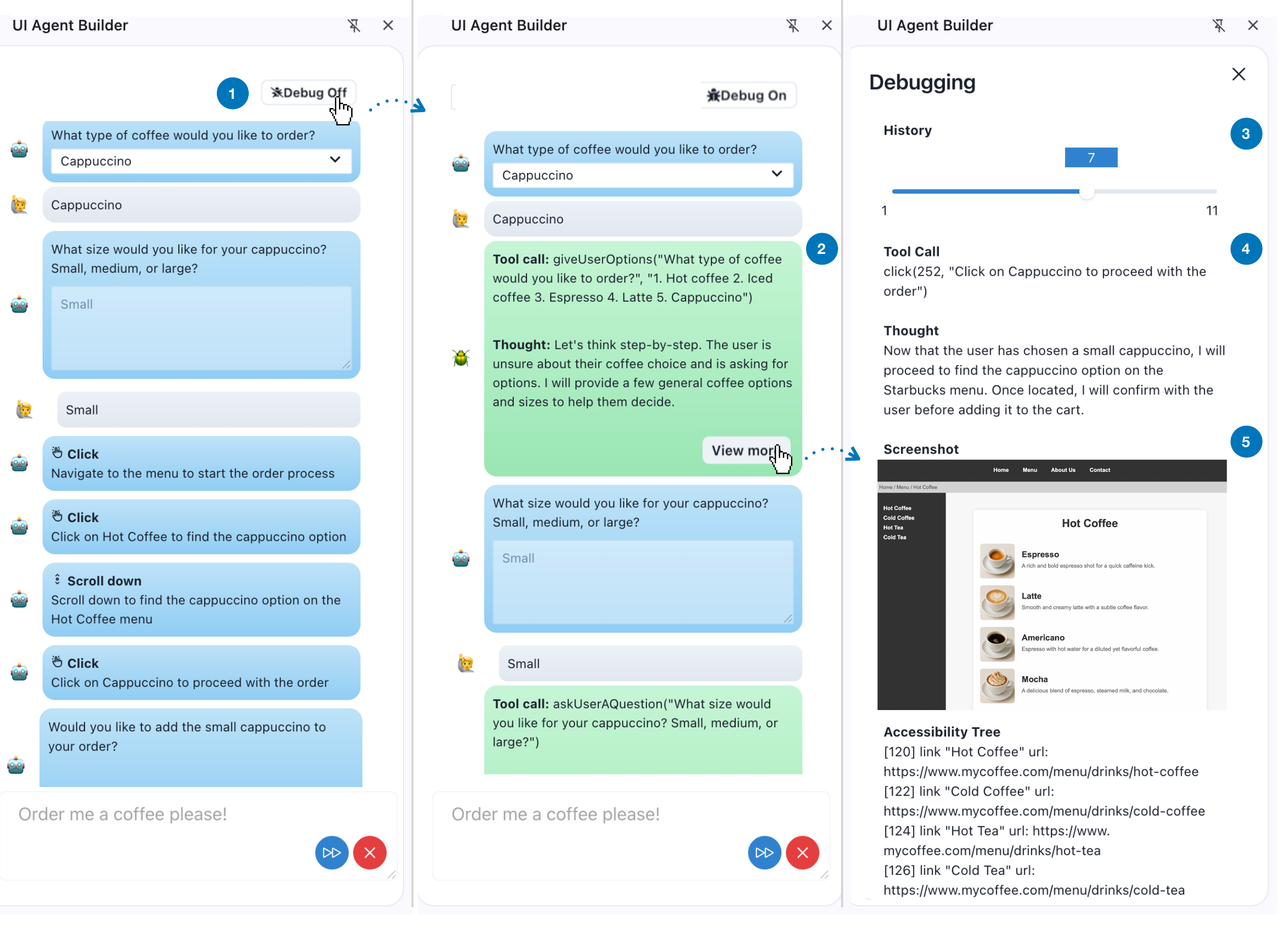} 
\caption{
Overview of the \tool~execution interface's {\sc{Debug Mode}}, which can be activated (middle, right) or deactivated (left) using the {\sc{Debug}} button \ilabel{1}.
During this mode, developers of agent prototypes can see the tool call and agent reasoning corresponding to each message in the chat \ilabel{1}.
Developers can also enter a {{\sc{Detailed Debugging}}} interface that includes the agent's inputs (i.e., webpage screenshot, webpage accessibility tree, text inputs) \ilabel{5} and outputs (i.e., tool call, agent reasoning) \ilabel{4}.
Developers of agent prototypes can navigate through the agent's previous actions using the slider \ilabel{3}.
}
\label{fig:execution-interface-debug}
\end{figure}

\scenario{To understand why the agent was unable to add the cappuccino to her cart (\textbf{A5}), Mei clicks on the {\sc{Debug}} button (\autoref{fig:execution-interface-debug}-1).
Scrubbing through the agent's previous actions (\autoref{fig:execution-interface-debug}-3), Mei realizes the correct UI element was not visible to the agent.
To fix this error, she writes in the {\sc{Other Instruction Prompt}}, \emph{``Scroll down the page if you are unable to perform a UI action after multiple tries, since the UI element may not be in view''}, which resolves the error on the next run.}

\subsubsection{\textbf{Debugging the agent}}
\label{sec:debug}
\tool~provides a {\sc{Debugging Mode}} (\autoref{fig:execution-interface-debug}) that exposes additional agent information such as tool calls and reasoning in green chat bubbles (\textbf{C6}).
A \textsc{Detailed Debug} view shows additional inputs to the agent, including the webpage screenshot, webpage accessibility tree, and full input context (\autoref{fig:execution-interface-debug}-4).
We included this interaction to explore the design consideration of \taxonomyCode{UI context}.
To navigate the agent's inputs and outputs of previous actions, developers of agent prototypes can use a slider (\autoref{fig:execution-interface-debug}-3).

\subsubsection{Implementation}
\label{sec:implementation}
\tool~is implemented as two TypeScript React applications. The {\sc{Prototyping}} interface is hosted as a web application, and the {\sc{Execution}} interface is a modified version of the CowPilot Chrome browser extension~\cite{huq-etal-2025-cowpilot}.

The {\sc{Prototyping}} interface uses the ReactFlow library for the graphical editor. To generate a prompt from the graph, the workflow is searched depth-first from the \texttt{Start} node to convert paths into a text representation. These are fed into \texttt{GPT-4o}, which expands this into a detailed description of the workflow. To propagate edits to the exported prompt back into the workflow graph, the updated prompt and JSON representation of the workflow are input to \texttt{GPT-4o}, and it is prompted to update the JSON, which re-instantiates the workflow.

The {\sc{Execution}} interface reuses CowPilot~\cite{huq-etal-2025-cowpilot}'s underlying agent and its agent controls. We added significant configurability to CowPilot---adding natural language descriptions of UI actions, making the UI action preview in the webpage configurable, and adding debugging agent outputs. We also added the Interaction Components, which are implemented as tool prompts.

\subsection{Step 3: Empirical User Study}
\label{sec:in-situ-study}

We used \tool~as a design probe in an empirical study where participants used \tool~to create prototypes of different agent UX. %
We used the insights from the study to validate our design requirements for agent UX prototyping systems (Section~\ref{sec:formative-study}) and further understand specific tooling needs developers of agent prototypes have with respect to the Desired Capabilities.

\subsubsection{Methodology}
\label{sec:user-study-methodology}
The user study protocol was approved by our institution's IRB.
During the user study, \numSecondStudyParticipants participants completed an agent UX prototyping exercise and were encouraged to think aloud during the process.
Sessions were conducted remotely using video conferencing software and lasted approximately 90 minutes.
Sessions were recorded and transcribed.
We obtained participant consent prior to all sessions.

\subsubsection*{Protocol}
Before the study, participants completed a brief demographic survey.
Sessions began with a 30-minute on-boarding session, which included defining agents with examples, a demo of a baseline agent which accepts a user's query and attempts to execute it without further interaction, and a tutorial on \tool.
Participants were asked to discuss the UX limitations of the baseline agent to encourage thinking about agent UX.

Next, participants were given 30 minutes for a prototyping exercise: to design an agent user experience using \tool~for navigating a real-world, popular coffee chain's website.
They were given a list of tasks that end-users of the website may want the agent to execute, including open-ended tasks (e.g., to discover menu options), specific tasks, and tasks with risks (e.g., placing order).
Participants were instructed to improve the baseline agent's UX, such as how and when the agent should interact with a user.
Participants ran their agent UX prototypes on the website logged in with a real user account with associated payment information.
Participants were encouraged to think aloud as they worked on designing the agent and reflected on the user experience of their prototype. We occasionally reminded participants about available features in the probe, following approaches from similar studies~\cite{zamfirescu2023johnny}.
Given time constraints in the sessions, the {\sc{Agent Capabilities Prompt}} and {\sc{User Information Prompt}} as part of \tool~were pre-populated.
Participants interacted with \tool~by remotely controlling the researcher's screen with the video conferencing software.
At the end of the study, the participant completed a 30-minute exit interview to reflect on their experience and any challenges faced in the process.

Following best practices in conducting user studies~\cite{ko2015practical}, we performed user study pilots with three pilot participants.
Through the study pilots, we successfully confirmed the study timing, improved the study protocol instruction wording, and verified that \tool's functionality could serve as a probe for agent UX prototyping.
Following the interview, the pilot participants provided feedback on the clarity of the instructions and the usability of \tool.
The protocol and tool were updated based on this feedback.

\subsubsection*{Participants}
Sampling from the list of interested individuals obtained during the requirements elicitation study (Section~\ref{sec:formative-study-methodology}), we selected a different set of \numSecondStudyParticipants participants (US-P1 to US-P14) spanning a range of experience with agents, generative AI, and prompt engineering.
Participants held a range of job titles, including software engineers ($N=4$), program and project managers ($N=3$), designers ($N=1$), product managers ($N=1$), ML engineers ($N=2$), and non-software engineering roles ($N=3$).
All participants were users of generative AI, with their usage ranging from a weekly basis ($N=1$) to once daily ($N=5$) or more than once daily ($N=8$).
Finally, study participants had a range of familiarity with agents, from never using or seeing a demo of an agent ($N=4$) to seeing demos of an agent ($N=8$) to using agents ($N=2$).
In accordance with the requirements elicitation study, we again intentionally sampled for a range of participant backgrounds to capture the perspectives of the stakeholders involved in AI UX design~\cite{yang2020re, dove2017ux, yang2018investigating, kayacik2019identifying, subramonyam2025prototyping, yang2025orbit}.

\subsubsection*{Analysis}
The author who conducted the interview studies analyzed the transcripts and recordings of the user study sessions
by close-coding on participant's prototyping \textit{Activities} from the requirements elicitation study (Section~\ref{sec:formative-study-results}) to validate the presence of the Activities.
To understand how the UX design space of computer use agents are explored (\textbf{RQ2}), the author performed provisional coding~\cite{saldana2009coding} using the set of \textit{Desired Capabilities} from the requirements elicitation study as well as open coding.

\subsubsection{Results}
\label{sec:user-study-results}
Overall, participants used \tool~to create a variety of agent UX approaches, ranging from a conversational design with the intention to foster user trust (e.g., US-P1), to an agent that only executed tasks. 
Consistent with prompt-based prototyping~\cite{liang2025prompts, zamfirescu2023johnny}, all participants iterated between the Design and Inspection phases of agent UX prototyping to explore, evaluate, and refine their designs.
For instance, US-P10 designed their first prototype to communicate \pQuote{maximal}{10} information about each UI action, but after execution found it was a poor experience as it produced \pQuote{way too much text}{10} in the {\sc{Chat}} panel.
Meanwhile, US-P1 designed their agent to ask for lots of user input since they \pQuote{prefer knowing exactly what the agent is going to do}{1}, but later felt the level of interaction \pQuote{is too much work for the user}{1}.

For each Activity and Desired Capability from Section \ref{sec:formative-study-results}, we observed that study participants engaged with the affordances of \tool~instantiating that activity and/or capability, validating our findings on how to design agent UX prototyping systems.
Below, we report the specific support needed for prototyping UX in computer use agents that supplements our understanding of each Desired Capability.
We first describe how participants interacted with \tool's features related to the capability and then discuss key observations.

\subsubsection*{\textbf{\capabilityOne{1}}}
All participants utilized \tool's no-code interfaces to prototype the agent's behavior, but relied on the notations to varying degrees.
Some participants used the {\sc{Workflow}} interface, whereas others relied on natural language by directly modifying the {\sc{Workflow Prompt}} or {\sc{Other Instructions Prompt}}.
For example, several participants first authored a preliminary {\sc{Workflow}} and then used the tool's {\sc{Generate Prompt}} feature to produce a prompt as a starting point, which they subsequently reused or refined.
Others, such as US-P10, adopted the opposite strategy of generating a workflow from a prompt and verifying that it aligned with the intended behavior.
We elaborate on additional insights on this Desired Capability below.

\paragraph{Using workflow notations affords better reasoning about agent behavior, but are hard to author}
The workflow notation provided helpful visualizations of agent behavior: \pQuote{[The workflow] had some value in the fact that I could see it visually...what's going on step by step}{12}. 
It allowed participants to trace the workflow to reason about the agent's actions (US-P1, US-P2, US-P4, US-P8, US-P10, US-P13, US-P14), helping them to identify conflicts in logic (US-P1, US-P4) and verify behavior before executing the agent (US-P13).
This was helpful in identifying agent errors: US-P8 found that their agent skipped a \texttt{Plan} step by comparing their workflow to the agent's messages in the chat.
In addition, US-P14, a product manager, mentioned how workflow notations reflected their work practice: \pQuote{When I see my engineers...we [whiteboard] with the same building blocks before...coding. This is very helpful because you now see the idea of the flow}{14}.

However, most participants encountered difficulties in using the workflow notation for authoring agent UX prototypes.
Even with a tutorial to educate participants on the agent's tools, a majority of participants expressed confusions on these concepts while working independently.
Participants struggled to accurately recall the tools at the agent's disposal (US-P1, US-P3, US-P7, US-P8, US-P9, US-P10, US-P11) or confused the capabilities of the agent's different tools (US-P1, US-P7, US-P10).
Viewing image previews of the tool's appearance in the chat interface was effective in disambiguating tools for many participants.

On the other hand, supplemental control with natural language provides more fine-grained control over agent behavior.
The flexibility of natural language prompts enabled participants to easily customize nuances in the agents' interactions with users, ranging from how it should handle user queries (e.g., \texttt{Ask user maximum of 2 clarifying questions} (US-P14), to the tone it should adopt with the user (e.g., being \texttt{energetic and kind} (US-P1)).

\subsubsection*{\textbf{\capabilityTwo{1}}}
Although the {\sc{Agent Capabilities Prompt}} and {\sc{User Information Prompt}} in our study (Section~\ref{sec:user-study-methodology}) were pre-defined, many participants noticed and reacted to the pre-set constraints and knowledge of the end-user during prototyping.
This came in the form of running the agent in scenarios that involve personalization such as \pQuote{order me my favorite drink}{4}, with the intention to understand how the agent may account for personal information about the user. 
This enabled participants to consider the design considerations of \taxonomyCode{risks} as well as the \taxonomyCode{user profile}.
Below, we share an additional insight on this Desired Capability.

\paragraph{Providing support for the agent's knowledge of the user promotes reflection on privacy and personalization}
By modifying the {\sc{User Information Prompt}}, some participants were able to find out the agent's knowledge regarding the end-user's personal information that the participant had previously overlooked.
This enabled several participants to reflect on how the agent handled sensitive user information and what about the experience should be improved. 
For example, after observing the agent automatically inserting the user's password to make a purchase, US-P9 said: \pQuote{The agent didn't ask [for the password]... It just entered it for me... Definitely not the most comfortable thing}{9}.

\subsubsection*{\textbf{\capabilityThree{1}}}
Most participants modified what information the agent displayed about UI actions in the chat and whether the agent revealed its UI action indicator on the webpage, indicating that the \taxonomyCode{visibility of agent activities} is an important factor to be considered in agent UX design.
Some participants designed their agents to reveal less information, such as US-P1, who intentionally designed their agent to not show UI action previews on the webpage or any UI action information in the chat to avoid information overload.
Other participants wanted to reveal more information during execution (e.g., showing description of actions or agent reasoning in the chat), with the hope that users could better trust the agent.
A key challenge in designing the agent UX was \pQuote{minimiz[ing] information, but just giving the right information}{6}. 
We elaborate on two additional insights on this Desired Capability below.

\paragraph{End users and prototypers of agents require different support to account for differing information needs}
Decisions on how to design the UI of the agent were guided both by participants' expectations of end-user behavior and by their own needs as developers of agent prototypes.
Participants envisioned that end users might prefer limited information about agent actions (e.g., only action names, no plans), yet as developers of agent prototypes, they expressed a desire for a more complete picture of how the agent would approach the task (US-P5, US-P6, US-P9, US-P10, US-P11, US-P12).
For example, US-P6's user-facing design displayed only action names in the chat.
When US-P6 asked the agent to recommend items with oat milk, the agent navigated through each menu category webpages one at a time. 
However, US-P6 was unable to infer the agent’s overall strategy from action names alone and as a result, was unable to evaluate whether the agent was functioning correctly.

For debugging purposes as a developer of agent prototypes, some participants enabled the UI action previews on the webpage for visibility of the agent's UI actions:
\pQuote{I always wanted show UI [actions]...When I saw that, it was really easy to understand, `OK, it's clicking this'}{2}.
Other participants (US-P12, US-P14) chose to reveal agent reasoning \pQuote{since for a developer it's pretty useful}{12}.
Reasoning of the agent was most helpful for a post-hoc understanding of agent actions, rather than while the agent was running.
US-P9 struggled to interpet the agent's UI actions as it was running on the browser:
\pBlockQuo{The reasoning...post-decision was helpful. In the moment, [the agent] is at the page. What is it doing? Is it thinking about what to get? Did it already pick and it's waiting for the page to load?}{9}
Taken together, this indicates that the agent UX design considerations of \taxonomyCode{transparency of agent reasoning} and \taxonomyCode{visibility of agent activities} vary by who is interacting with the agent.

\paragraph{Support is needed for more fine-grained control over how the agent interacts with users}
Even with no-code interfaces, participants still wanted more support to understand what about the agent could be modified.
Several participants struggled to figure out the best way to modify the agent's tool outputs in the chat (US-P6, US-P8, US-P9).
As a result, natural language and the workflow notation seemed insufficient for prototyping agents: 
\pBlockQuo{They both feel dumbed down... 
I feel like my agency has been taken...[when the agent] doesn't work and then I get frustrated, but I only know these are the only options available.}{8}

Participants requested more granular supports to modify the agent's behavior (US-P6, US-P8, US-P9, US-P14) for additional control over the \taxonomyCode{transparency of agent reasoning} and \taxonomyCode{description of agent actions}.
Requested features included mechanisms to change the number of options that the agent provided to the user in \texttt{Interact} nodes or scaffolds to reduce the wordiness of agent's action descriptions and reasoning for \texttt{UI Actions} nodes.

\subsubsection*{\textbf{C4. Provide components that enable the agent to invoke different user interactions.}}
When designing the agent, participants explored the interaction components available to the agent in the {\sc{Workflow}}, which allowed them to understand the \pQuote{capabilities of this entire agent}{12} and \pQuote{knowing what actions it can make}{13}.
Below, we discuss one additional insight on this Desired Capability.

\paragraph{Providing components helps scaffold the agent design space}
Showing the tool options available on the {\sc{Workflow}} interface clearly defined the design space for participants, as it helped participants understand \pQuote{the limited amount of tools...in place that I could use for the agent}{5}.
This was helpful even for developing natural language prompts for the agent: \pQuote{[The {\sc{Workflow}}] helped frame a lot of the different types of instructions... I got the `Provide options' terminology straight from the builder.}{9}

Without relying on any scaffolding, participants found it challenging to compose a prompt to define granular specifications of the agent's behavior in natural language (US-P2, US-P5, US-P9, US-P12, US-P13), such as invoking specific tools.
For example, US-P12 prompted the agent to ask the user whether they would like to order anything else once an item was added to the cart, with the intention of having the agent show dropdown menu options.
While the agent was able to send the appropriate message in the chat, it used the incorrect tool, and as a result did not provide the expected UI component the participant requested.

\subsubsection*{\textbf{C5. Provide an environment to run and control the agent.}}
All participants ran and observed their agent at least once to understand the user experience with the agent.
A majority of participants used the {\sc{Stop}} or {\sc{Pause}} controls to intervene while the agent was running to focus on interesting behavior.
This usually occurred when the participant realized that the agent encountered an error and needed to read the agent's reasoning that led to the error (US-P3, US-P4, US-P5, US-P6, US-P7, US-P8, US-P12, US-P13).
Only US-P9 used the {\sc{Pause}} and {\sc{Resume}} features for Cowpilot~\cite{huq-etal-2025-cowpilot}'s intended purpose, which was was to guide the agent to a better set of actions to complete the given task~\cite{huq-etal-2025-cowpilot}.
We further discuss one insight regarding this Desired Capability below.

\paragraph{Support is needed to communicate what actions the agent is performing in the current run}
Although participants were very attentive to the agent's actions in the webpage and chat panel as the agent was running, many still lost track about whether and when the agent was running, encountering an error, stopped, or waiting for user input (US-P1, US-P3, US-P4, US-P5, US-P7, US-P9, US-P11).
This challenge was worsened when the participant designed the agent to not display action descriptions and/or reasoning in the chat or chose to hide the agent's actions on the webpage.
For example, because US-P1 did not enable the UI action previews in the webpage, they failed to realize their agent added four croissants to the cart.

In addition, each time the agent took an action, participants needed to consider both the webpage the agent was running on and the chat panel that contained the agent's messages and its action reasoning, 
creating difficulties in processing this information in real time (US-P3, US-P8, US-P9, US-P10, US-P12, US-P13, US-P14): \pQuote{The most difficult part was just trying to keep up with what's going on the left [webpage] and what...the agent is thinking}{12}.
As a result, many participants missed quick and subtle actions (e.g., adding items to the cart) and thus were unable to accurately recall the agent's actions (US-P1, US-P3, US-P4, US-P6, US-P7, US-P10, US-P11).
For example, while proceeding to checkout, US-P3 was surprised to find that four items were in the cart, even though they expected one.

To handle the volume of information, some participants had to pause the agent to process at their own pace (US-P3, US-P4, US-P5, US-P6, US-P7, US-P8, US-P14).
Others had to focus on one source of runtime information while ignoring others. 
US-P10 chose to not read the chat and only monitored the webpage.

\subsubsection*{\textbf{C6. Help user debug the agent's runtime behavior}}
A majority of participants used \tool's {\sc{Debug Mode}} to understand the agent's runtime behavior.
Most viewed only the agent's reasoning (US-P1, US-P3, US-P4, US-P5, US-P6, US-P7, US-P9, US-P10, US-P11, US-P12, US-P13), while several others further investigated the agent's action inputs in the {\sc{Detailed Debugging}} interface (US-P4, US-P5, US-P6, US-P7, US-P8, US-P9).
We elaborate on additional tooling support for this Desired Capability below.

\paragraph{Support is needed to explore the input context and tools provided to the agent}
Many participants found the agent input context and outputs in the {\sc{Debug Mode}} (e.g., webpage accessibility tree, text input from a given timestep) to be difficult to interpret (US-P4, US-P5, US-P6, US-P7, US-P9, US-P10).
As a result, participants struggled to debug the agent's behavior because they misinterpreted or did not understand how the agent made decisions.
For example, US-P10 mistook the accessibility tree text as agent reasoning and incorrectly recounted the agent's previous actions.

In addition, most participants relied mainly on the reasoning of the agent.
However, the reasoning had issues like hallucinations (US-P1, US-P11) or insufficient detail (US-P6).
For example, when US-P1 asked their agent to order a large latte into their cart, the agent in actuality added a medium but said in the reasoning it added a large.
Thus, \taxonomyCode{agent errors} that are difficult to discover can hinder debugging of agents.
We discuss tooling support for the Desired Capability below.

\paragraph{Scaffolds are needed to help build accurate mental models of the agent's behavior.}
Participants were unable to understand across the tool prompts and input context to identify why unwanted behavior might occur. 
For example, several participants encountered the issue where only a small number of items appeared in the drop-down menu tool, but only one participant (US-P6) was able to figure out that it was due to constraints on the agent's current viewport and the tool prompt.

\mybox{
\faArrowCircleRight\xspace\textbf{Summary of key findings (RQ2):}
Agent UX prototyping requires support for the following six activities: designing the scope and boundaries (\textbf{A1}), information display (\textbf{A2}), and user interactions (\textbf{A3}) of the agent; running the agent UX prototype (\textbf{A4}); and understanding the agent's runtime behavior (\textbf{A5}).
Agent UX prototyping systems should  use no-code interfaces (\textbf{C1}), constrain the agent's task space and knowledge of the user (\textbf{C2}), define the UI of the agent (\textbf{C3}), provide components to invoke different user interactions (\textbf{C4}), provide an environment to execute the agent (\textbf{C5}), and debug the agent's runtime behavior (\textbf{C6}).
}
\section{Discussion}
In this work, we explore how to design the UX of computer use agents.
First, we outline the UX design space of computer use agents in a design consideration taxonomy (\textbf{RQ1}, Section~\ref{sec:taxonomy}).
Given the importance of prototyping in the role of designing AI UX~\cite{yang2020re}, we then translated the taxonomy into a design probe to understand what support is needed by developers of agent prototypes to prototype different agent UX approaches (\textbf{RQ2}, Section~\ref{sec:design-probe}).
Informed by six Desired Capabilities in agent UX prototyping systems identified in a requirements elicitation study (Section~\ref{sec:formative-study}), we developed a design probe, \tool, a system that helps developers of agent prototypes explore the UX design space of agents through prototyping.
By instantiating the Desired Capabilities and taxonomy in \tool~and observing its use in an empirical user study (Section~\ref{sec:in-situ-study}), we provide insights on the support required for prototyping agent UX.
Below, we synthesize the findings across all the studies.
We discuss the implications for computer use agents (Section~\ref{sec:implications-cua}) and for developing agent UX (Section~\ref{sec:implications-agent-ux}). 
We then conclude with the limitations of this work (Section~\ref{sec:limitations}).

\subsection{Implications for Computer Use Agents}
\label{sec:implications-cua}
We anticipate that the design dimensions identified in the taxonomy will provide a shared vocabulary and useful directions to help designers reason about factors related to the UX of computer use agents.
Although developed for computer use agents, future work should understand if they may be applicable to general agentic systems.

Below, we compare the taxonomy of UX design considerations with those from the early interface agent literature (Section~\ref{sec:cua-vs-interface-agents}) and conclude with a discussion of key challenges and open opportunities for the UX of computer use agents (Section~\ref{sec:cua-challenges-opportunities}).

\subsubsection{Building on the design principles in early interface agents}
\label{sec:cua-vs-interface-agents}
We find that three of the four main categories identified in the taxonomy of UX considerations of agents largely aligns with those identified in early work with interface agents.
However, computer use agents also introduce new UX design considerations, as the only category less directly addressed by this literature is the \emph{user mental model \& expectations}, reflecting the difficulty of helping users develop a detailed and accurate mental model of agents~\cite{bansal2024challenges, brachman2025building} and AI at large~\cite{tenhundfeld2021my, anderson2020mental, horstmann2023alexa}.

In terms of similarity, for the \emph{user query}, both types of agents include UX considerations of the \taxonomyCode{user profile}, like remembering recent interactions~\cite{horvitz1999principle} and information the user provides; handling \taxonomyCode{ambiguity}~\cite{horvitz1999principle} in the user's goals;  \taxonomyCode{contextual factors}, such as the user's attention~\cite{lieberman1997autonomous, horvitz1999principle}; and \taxonomyCode{when the user enters the query} to resolve ambiguity~\cite{horvitz1999principle}.
For the \emph{explainability of agent activities}, the early work on interface agents emphasized the importance of helping users understand and trust agents~\cite{shneiderman1997direct}.
This literature discussed the \taxonomyCode{visibility of agent activities}, such as how the agent should appear to the user in the UI~\cite{shneiderman1997direct, dehn2000impact} (e.g., whether the agent was visible) and techniques for the \taxonomyCode{communication of runtime status}~\cite{lieberman1997autonomous, lieberman1995Letizia}.
In terms of \emph{user control}, the literature on interface agents debated whether agents should be used in \taxonomyCode{high impact scenarios}~\cite{lieberman1997autonomous}, recognizing the cost of \taxonomyCode{errors} to the agent's utility~\cite{horvitz1999principle}.
Command over interface agents was also essential for users to never feel out of control~\cite{shneiderman1997direct}; thus, allowing for \taxonomyCode{user intervention during agent execution} via agent controls (e.g., direct invocation and dismissal) was a recommended practice.

\subsubsection{Challenges \& opportunities for computer use agents}
\label{sec:cua-challenges-opportunities}
In spite of the similarities with interface agents, modern computer use agents introduce new UX design considerations.
While some new considerations are specific to the reasoning and planning nature of (M)LLM-based agents (e.g., \taxonomyCode{presentation of plan}, \taxonomyCode{transparency of agent reasoning}), many of them are also fueled by the increased intelligence and autonomy unique to generative AI.
For instance, an (M)LLM's ability to handle arbitrary text and images introduces new design considerations of \taxonomyCode{levels of expression} and \taxonomyCode{modality of user input}.
These improved capabilities place increasing importance on ensuring user \taxonomyCode{safety} and mitigating \taxonomyCode{risks} of using computer use agents, which is an emerging field of research.
While existing work have outlined some risks~\cite{zhang2025interaction}, further research is required to develop mechanisms to mitigate the harms of computer use agents, such as how to appropriately set the \taxonomyCode{scope of the agent}.
Future work could also investigate how to support the development of the \emph{user mental model \& expectations} of the agent, as this was the only category of UX design considerations not deeply discussed in the literature of interface agents.
The user's mental model of agents is an open challenge in human-agent interaction~\cite{bansal2024challenges, brachman2025building}; future research is needed on how to communicate the \taxonomyCode{agent capability} to users.

\subsection{Implications for Prototyping Agent UX}
\label{sec:implications-agent-ux}
Based on the findings in this work, we discuss implications of this work on prototyping agent UX, with an eye towards design recommendations for future tools that prototype agent UX.
Although this work primarily focuses on the UX of computer use agents, we believe these recommendations can also apply to the development of agents that interface with users (e.g., in-IDE coding agents), given the similar nature of the problem space.

\subsubsection{Use workflow notations to visualize agent behavior, while additional design supports are needed.}
Node-based workflow notations are a commonly utilized approach for systems that support AI development~\cite{esposito2023end}.
In fact, many current state-of-the-art agent development tools in research (e.g., FlowForge~\cite{hao2025flowforge}, ChainForge~\cite{arawjo2024chainforge}, PromptChainer~\cite{wu2022promptchainer}) and in practice (e.g., Zapier~\cite{zapier2026zapier}, n8n~\cite{n8n2026n8n}, and OpenAI's AgentBuilder~\cite{openai2026agentbuilder}), rely on node-based notations to build agentic workflows and other LLM pipelines.
However, our study results suggest that workflow notations may not be the most effective for designing and constructing agent workflows.
In the empirical user study of \tool, numerous participants struggled to learn how to express what they wanted in \tool's workflow notation.
This reflects how notations designed for end-users---like the workflow notation---can be challenging to use, as it is difficult to design programs, learn the notation, and write in it~\cite{ko2004six}.
Additional research is necessary to study new abstractions or scaffolds for designing agents.
Additionally, in the empirical user study of \tool, some participants did use the workflow as a visualization of the intended workflow, which was helpful during debugging.
This suggests that the workflow notation could be better suited as a visualization of agent behavior rather than a mechanism for agent authoring, which could be explored further in future work.

In addition, the wide range of possible capabilities and outputs of generative AI make it particularly challenging as a design medium~\cite{yang2020re}.
As a result, the design space of agents is vast, which poses challenges in agent development~\cite{hao2025flowforge, anonymous2026redacted}.
This was echoed by study participants, who expressed confusion in knowing what to include in the specification of agent behavior while using open-ended mediums like natural language (US-P2, US-P5, US-P9, US-P12, US-P13).
Thus, our results point to the workflow notation being a helpful scaffold to design agents, as it is a visual representation of the UX design space and can reveal the relationships between different design considerations.
For example, while using \tool, multiple study participants referred to the agent tools as a natural scaffold to constrain the agent design space.
This corroborates prior work, which suggests that workflow notation scaffolds for developers' design thinking improves the developer experience, reduces development time, and produces more diverse agent designs~\cite{hao2025flowforge}.

\subsubsection{List which tools the agent has access to scope the agent design space}
Our findings underscore that an individual's mental model of the agent is vital while prototyping, as underscored in prior work~\cite{brachman2025building, bansal2024challenges}.
In the empirical user study of \tool, we observe that developing accurate mental models of agents is challenging when the individual who is prototyping lacks an understanding of the agent's implementation, particularly in what tools it has access to.
This corroborates~\citet{brachman2025building}, who finds that individuals interacting with agents often lack insight on the inner workings of agents.
Without a well-formed understanding of the agent's tools, study participants struggled to debug and modify agent behavior.
Therefore, providing interfaces to explore and modify agent tools is essential for agent development.
For designing agent UX, these tools should also consider when to invoke inputs from the users via \formativeCode{interaction components}, since agent UX prototyping systems should provide components for agents to invoke different user interactions (\textbf{C4}).

\subsubsection{Support testing of an agent with different user personas}
From the requirements elicitation study, we find that running the agent (\textbf{A4}) is an important step of prototyping agents.
The results from the empirical user study of \tool~further elucidate that while running the agent, the information needs of agent users and developers of agent prototypes differ and sometimes even conflict.
We observe that as a developer of agent UX prototypes, study participants required more information about the \emph{explainability of agent activities}, which is necessary for agent debugging~\cite{epperson2025interactive}.
At the same time, participants felt that providing this level of detail on agent activities could detract from the agent UX~\cite{anonymous2026redacted}.
Therefore, future agent UX prototyping systems support the testing of different personas that may interact with the agents.
Agent interactions with a developer of agent prototypes should support detailed logs of agent \formativeCode{text inputs and outputs} as well as additional context such as \formativeCode{UI state}, whereas the  testing for a user persona should have more flexibility in what information is displayed.
Additional personas to test the agent UX on include users with varying personal preferences in their \taxonomyCode{user profile}, as well as types of goals within the \taxonomyCode{contextual factors} (e.g., having exploratory, open-ended queries versus executing specific commands), which we find are important considerations in related work~\cite{anonymous2026redacted}.

\subsubsection{Provide direct manipulation interfaces to control agent execution}
Design guidelines of AI emphasize the importance of global controls of the AI~\cite{amershi2019guidelines}, which we find is necessary for understanding the agent's runtime behavior (\textbf{A5}) in the form of \formativeCode{agent controls}.
Prior work has suggested controls for pausing, stopping, and undoing agent actions~\cite{liang2025tabletalk, huq-etal-2025-cowpilot, epperson2025interactive}.
Our empirical study with \tool~suggests that in addition to these controls, additional support is necessary, which could be explored by future research.
This includes features that support processing the agent's outputs, such as slowing down execution speed, replaying agent actions, and annotating agent execution history.
Research indicates that interactive debugging features, such as changing agent messages mid-execution, are helpful for agent developers~\cite{epperson2025interactive}.
Without sufficient interfaces to control and make sense of agent execution, study participants relied on repeatedly pausing and resuming the agent's actions to slow down the agent (US-P3, US-P4, US-P5, US-P6, US-P7, US-P8, US-P14) or focusing only on a subset of the agent's outputs (US-P10).

\subsection{Limitations}
\label{sec:limitations}
Below, we discuss the limitations of \tool, the taxonomy study, requirements elicitation study, and empirical user study.
Given the exploratory and qualitative nature of the work, the studies have several limitations.
In terms of external validity, participants were recruited within the same technology company for all studies.
This could introduce demographic and professional bias towards tech-savvy, well-educated, English-speaking populations, potentially limiting the generalizability of results.
In addition, the taxonomy was based of an initial sampling of 11 real-world agents, which may not be representative of all computer use agents.
Finally, participants of the empirical user study worked under time pressure on a single task, which may not be representative of all working conditions and tasks.
Future work could explore the effects of \tool~on agent UX prototyping with more complex and realistic tasks, such as through a longitudinal field study.

In terms of internal validity, the results from the requirements elicitation study could be impacted by memory bias.
We reduced this threat by having participants ground their reflections on UX design considerations based on a specific experience with a computer use agent.
In addition, \tool---used during the empirical user study---relied on LLMs to generate the workflow visualization, which are known to hallucinate or produce incorrect answers.
The main aim of the empirical user study was to elicit observations on the support required for agent UX prototyping, leaving the question of the accuracy of the generation of the workflow visualization unanswered.
To reduce this threat, we piloted the empirical user study protocol with three participants to ensure the tool's usability and its ability to function as a design probe.
In addition, during the study itself, we did not observe any adverse effects on participants' interactions with the tool or their ability to prototype agents.
However, the overall quality of the workflow generations remain an open question and could be more systematically evaluated in future work.

Given the limitation of the studies, the results of this work should be treated as hypotheses to be validated in future work, such as via controlled studies.
For \textbf{RQ1}, we do not claim that the taxonomy is comprehensive of all UX design considerations of computer use agents and important dimensions may not be represented.
For example, an important dimension not captured in our taxonomy is accessibility.
Although it was beyond the scope of this project, we recognize its significance, particularly given the potential of using AI-powered computer use agents for accessibility features \cite{vu2023voicify}.
Future research could extend the taxonomy to include accessibility, for example, by exploring how it applies to the needs of visually impaired users when interacting with agents.
For \textbf{RQ2}, the insights derived from the empirical user study of \tool~may not generalize to the full design space captured by the taxonomy.
This is because \tool~instantiates only web agents, whereas the taxonomy spans desktop, browser, and mobile environments.
As a result, prototyping needs for other environments may not be surfaced from our probe, which could be investigated by future work.

\section{Conclusion}
In this work, we explore the design space of computer use agent UX (\textbf{RQ1}) as well as scaffolds that are needed to support designing agent UX via prototyping (\textbf{RQ2}).
To answer \textbf{RQ1}, we develop a taxonomy of agent UX design considerations by reviewing eleven modern computer use agents as well as interviewing eight AI and UX practitioners.
Next, we translate the taxonomy into an agent UX prototyping design probe to answer \textbf{RQ2} in a three-part study.
First, we first outline the design requirements of agent UX prototyping systems based on interviews with \numFirstStudyParticipants individuals who have a variety of prior experiences with agents to represent the diverse backgrounds that are involved in AI design.
The design requirements include an enumeration of five Activities in the agent UX prototyping process, from which we derive six Desired Capabilities that agent UX prototyping systems should support.
Next, we use these design requirements to inform the design of \tool, a design probe that enables the design, development, and execution of different agent UX prototypes.
We use \tool~as a probe to conduct an empirical user study of agent UX prototyping with \numSecondStudyParticipants participants. 
Through this study, we validate the design requirements and enumerate the support that is needed for agent UX prototyping.
Based on these findings, we conclude with a set of design recommendations for agent UX prototyping systems.
We provide supplemental materials for this work~\cite{supplemental-materials}.
\ifanonymous
\else
We publicly release the source code for \tool: \url{https://github.com/apple/ml-agentuxlab}.

\bibliographystyle{ACM-Reference-Format}
\bibliography{sample-base}

\newpage
\appendix
\section*{Appendix}
We include a subset of the supplemental materials for this work.
For the complete set of supplementary materials, including the protocols, codebooks, and video figure, please refer to the materials on ScholarOne~\cite{supplemental-materials}. 
We will make all materials public upon acceptance of the work.

\section{Design Probe: \tool}
Below, we describe how \tool~can support the prototyping of different agent experiences by illustrating three distinct agent UX prototypes for the same user query (Section~\ref{sec:agent-prototype-examples}).

\subsection{Agent UX Prototype Examples}
\label{sec:agent-prototype-examples}
\tool~supports the construction of different types of agent UX prototypes that can vary in the level of interaction with the user and the amount of information the agent communicates about its UI actions while it is executing.
We illustrate a range of different agent experiences that \tool~can prototype in Figure~\ref{fig:example-prototypes} and describe it further below.
Each prototype helps the user order a cappuccino with the ambiguous user query, "Order me a coffee please!"

Prototype 1 is an agent that is interactive by clarifying the order and asking for confirmation before adding to the cart.
It also communicates information about each UI action it performs on the screen.
Next, Prototype 2 is an agent that approaches the same problem without any user interaction.
Instead, it shows the user a plan of how it will complete the actions, summarizes its actions in messages, and makes decisions about what is ordered without user input.
To reduce the amount of information shown to the user, this agent only shows the UI action's name each time it performs an action on the webpage.
Finally, Prototype 3 is an agent that is designed to be friendly, communicative, and interactive for the user query by clarifying the order and asking for confirmation before adding to the cart.
This agent does not show any information about each UI action it performs on the screen to reduce the amount of text in the chat.

To highlight the differences between agent UX prototypes, compared to Prototypes 1 and 3, Prototype 2 has much less user interaction.
Compared to Prototypes 1 and 2, Prototype 3 does not communicate any information about the UI actions; further, it is more conversational than Prototype 1.
We now describe how to construct each agent UX prototype in \tool~below.

\begin{figure}[t!]
\centering
\includegraphics[trim=0 0 0 0, clip, width=0.95\linewidth, keepaspectratio, page=5]{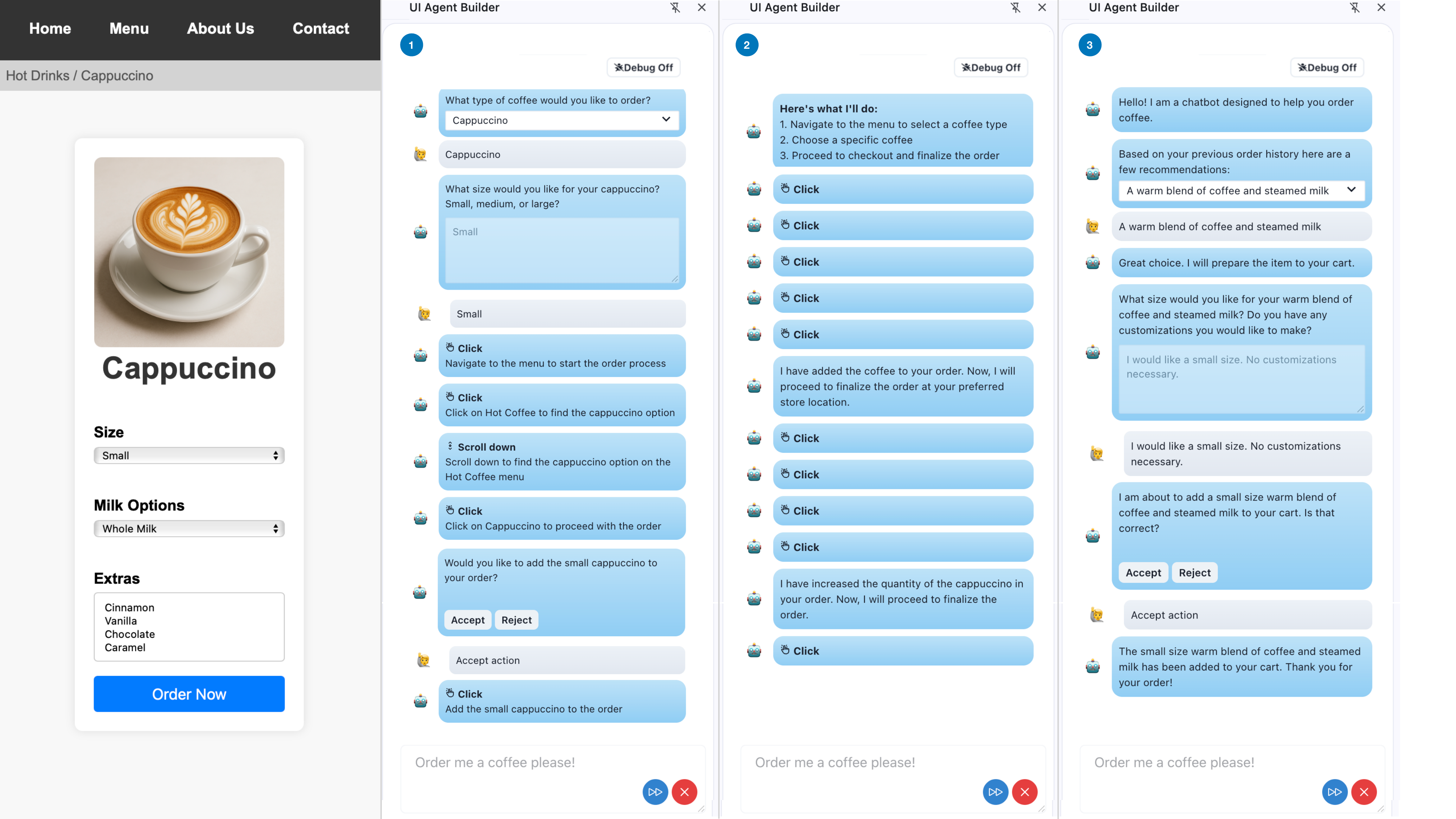} 
\caption{
Example agent UX prototypes \ilabel{1} \ilabel{2} \ilabel{3} for the same task and user query, which is to order a cappuccino with the user query "Order me a coffee please!" Prototype \ilabel{1} is interactive, and shows the most information about the agent's UI actions; prototype \ilabel{2} executes the task autonomously, and shows some information about the agent's UI actions; and prototype \ilabel{3} is interactive, but shows no information abotu the agent's UI actions.
}
\label{fig:example-prototypes}
\end{figure}

\subsubsection{Prototype 1}
Prototype 1 is the interactive agent that clarifies coffee orders, confirms with users, and communicates information about UI actions.
This agent can be created in \tool~by adding in user interaction with \texttt{Confirm} and \texttt{Interact} nodes. 
To build this agent in \tool, the developer can create a flow to \texttt{Interact} with the user, perform \texttt{UI Actions}, and \texttt{Confirm} with the user before adding to the cart (i.e., \texttt{Start} $\rightarrow$ \texttt{Interact} $\rightarrow$ \texttt{UI Actions} $\rightarrow_{if\_add\_cart}$ \texttt{Confirm} $\rightarrow$ \texttt{End}).
The developer can also set the \texttt{UI Actions} to show the UI action's name (e.g., scroll, click) and a brief description of the UI action.

\subsubsection{Prototype 2}
Prototype 2 is not an interactive experience. 
It instead shows a plan of actions, summarizes actions in messages, and only shows the UI action's name.
To prototype this agent in \tool, the developer can create a flow that displays a \texttt{Plan} to the user and sends a \texttt{Message} when major steps in the plan are completed (i.e., \texttt{Start} $\rightarrow$ \texttt{Plan} $\rightarrow$ \texttt{UI Actions} $\rightarrow_{if\_step\_done}$ \texttt{Message} $\rightarrow$ \texttt{End}).
The developer can set the \texttt{UI Actions} to show the UI action's name only.

\subsubsection{Prototype 3}
Prototype 3 is a friendly, communicative, and interactive agent that clarifies orders, asks for confirmation, and sends update messages after important steps are accomplished.
It shows no information about each UI action to reduce the amount of text in the chat.
To build this agent in the tool, the developer can create a flow to send a welcome \texttt{Message}, \texttt{Interact} with the user to understand their order, \texttt{Message} the user a summary of the actions, and \texttt{Confirm} with the user before adding an item to the cart (i.e., \texttt{Start} $\rightarrow_{welcome\_message}$ \texttt{Message} $\rightarrow$ \texttt{Interact} $\rightarrow_{summarize\_order}$ \texttt{Message} $\rightarrow$ \texttt{UI Actions} $\rightarrow_{if\_add\_cart}$ \texttt{Confirm} $\rightarrow$ \texttt{End}).
The developer can show no information about the agent's \texttt{UI Actions} in the chat and in the webpage.

\section{Empirical User Study of \tool}
We discuss the types of examples participants created with \tool~(Appendix~\ref{sec:in-situ-study-prototype-examples}) as well as the instructions given to study participants for the agent UX prototyping exercise (Appendix~\ref{sec:in-situ-study-instructions}).

\subsection{Participant Agent UX Prototype Examples}
\label{sec:in-situ-study-prototype-examples}
Below, we illustrate the range of agent experiences participants were able to prototype using \tool.

\begin{figure}[t!]
\centering
\includegraphics[trim=0 25 1100 0, clip, width=0.75\linewidth, keepaspectratio, page=1]{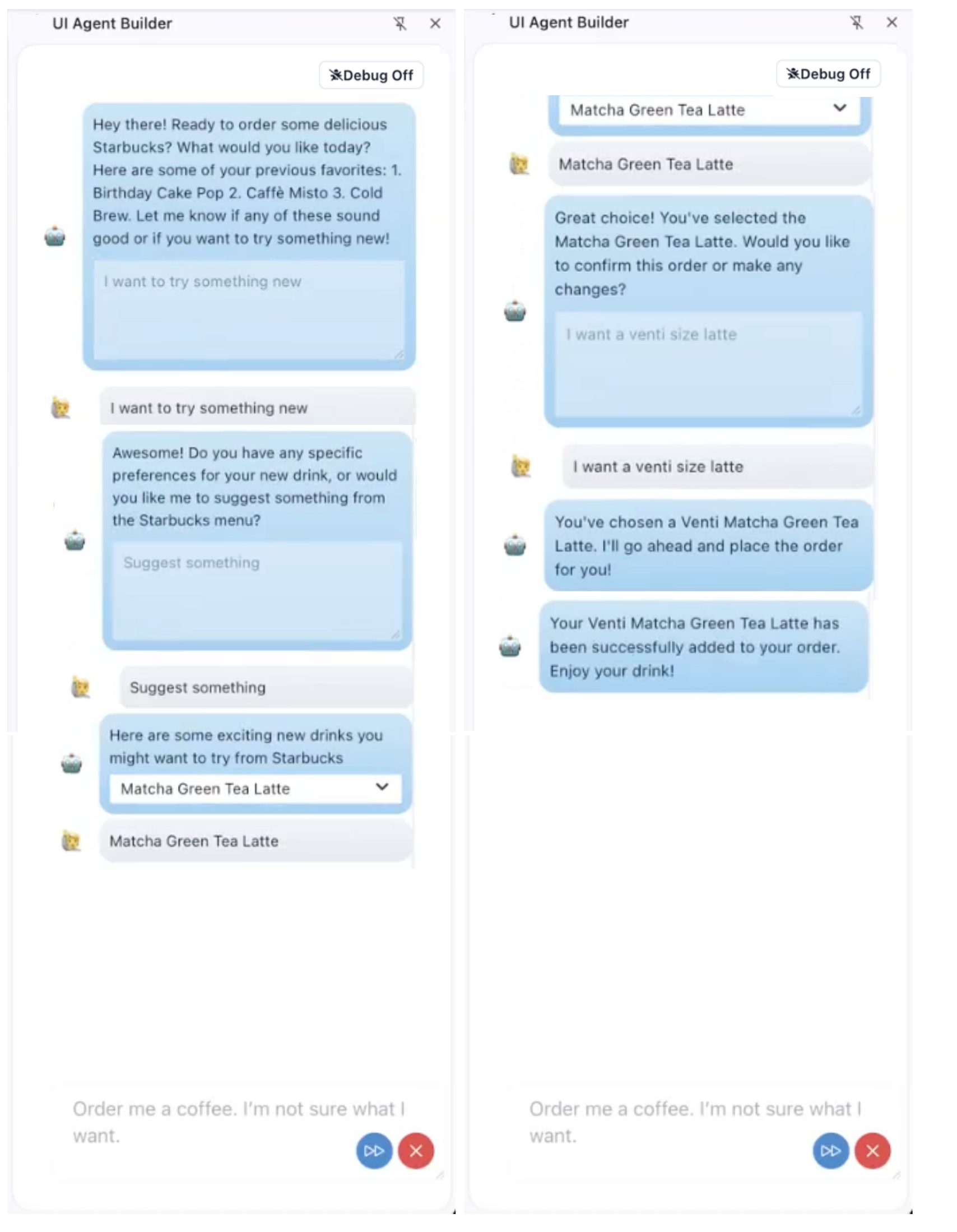} 
\caption{
The agent UX prototype that P1 created during the \emph{in situ} agent UX prototyping study for the input \pQuote{Order me a coffee. I'm not sure what I want}{1}.
}
\label{fig:p1-example-prototype}
\end{figure}

\subsubsection{P1 agent UX prototypes}
P1's agent is shown in Figure~\ref{fig:p1-example-prototype}, where they inputted \pQuote{Order me a coffee. I'm not sure what I want}{1}, and worked with the agent to order a matcha green tea latte.
P1 designed their agent to be friendly, \texttt{energetic} (P1), and interactive.
P1's workflow was designed to have user interaction and included two \texttt{Interact} nodes and a \texttt{Confirm} node.
P1 designed their agent to show no information (i.e., no action name, action description, or agent reasoning) about the agent's UI actions in the chat.
As a result, while executing the agent, P1's agent asked for user input four times to clarify and confirm the user's order before proceeding to complete the order.

\begin{figure}[t!]
\centering
\includegraphics[trim=0 25 1500 0, clip, width=0.37\linewidth, keepaspectratio, page=2]{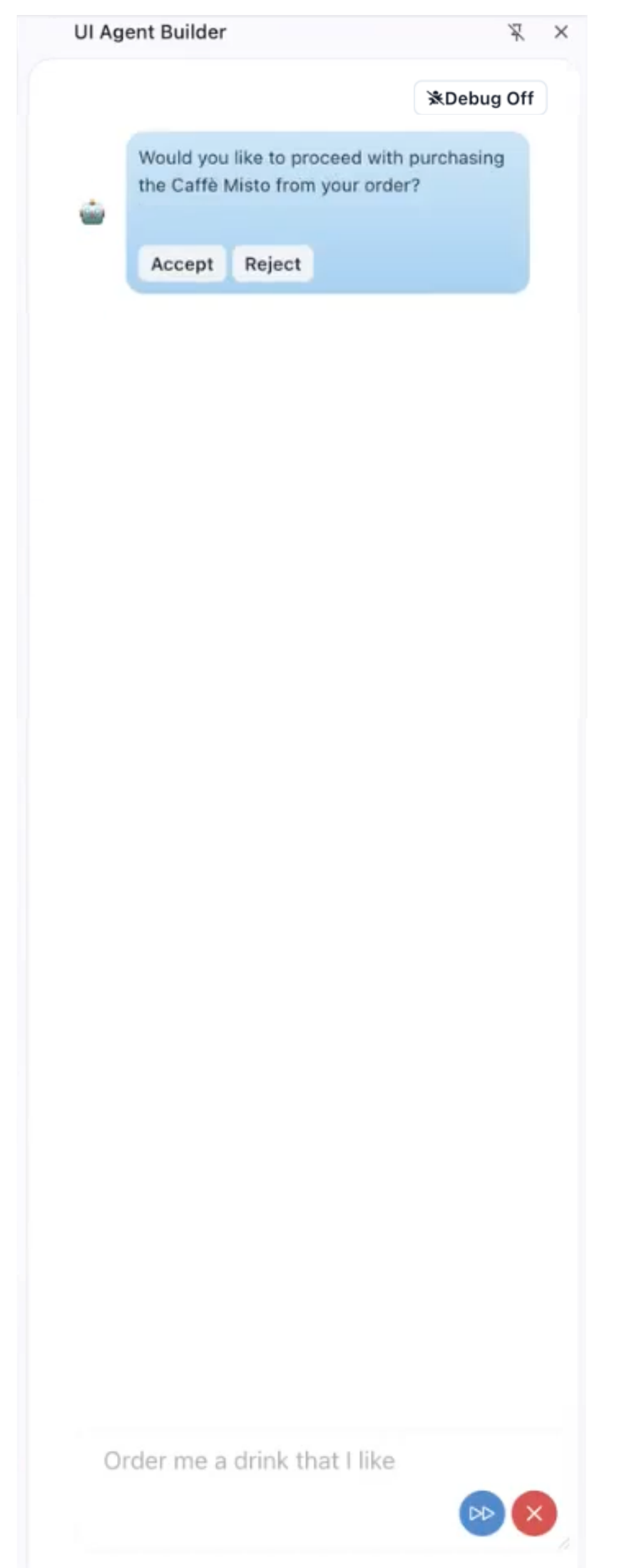} 
\caption{
The agent UX prototype that P4 created during the \emph{in situ} agent UX prototyping study for the input \pQuote{Order me a drink that I like}{4}.
}
\label{fig:p4-example-prototype}
\end{figure}

\subsubsection{P4 agent UX prototypes}
P4's agent is shown in Figure~\ref{fig:p4-example-prototype}, where P4 gave the query \pQuote{Order me a drink that I like}{4}, and worked with the agent to order a drink called Caffè Misto.
P4's agent workflow was \texttt{Start} $\rightarrow$ \texttt{UI Actions} $\rightarrow_{high\_risk\_action}$ $\rightarrow$ \texttt{End}.
This agent also did not show any information about UI actions (i.e., action name, action description, agent reasoning) in the chat.
Therefore, when P4's agent UX prototype was run, the chat pane was empty with the exception of a single confirmation message before purchasing the Caffè Misto, which was deemed by the agent as a potentially risky action.

\begin{figure}[t!]
\centering
\includegraphics[trim=0 25 1500 0, clip, width=0.37\linewidth, keepaspectratio, page=4]{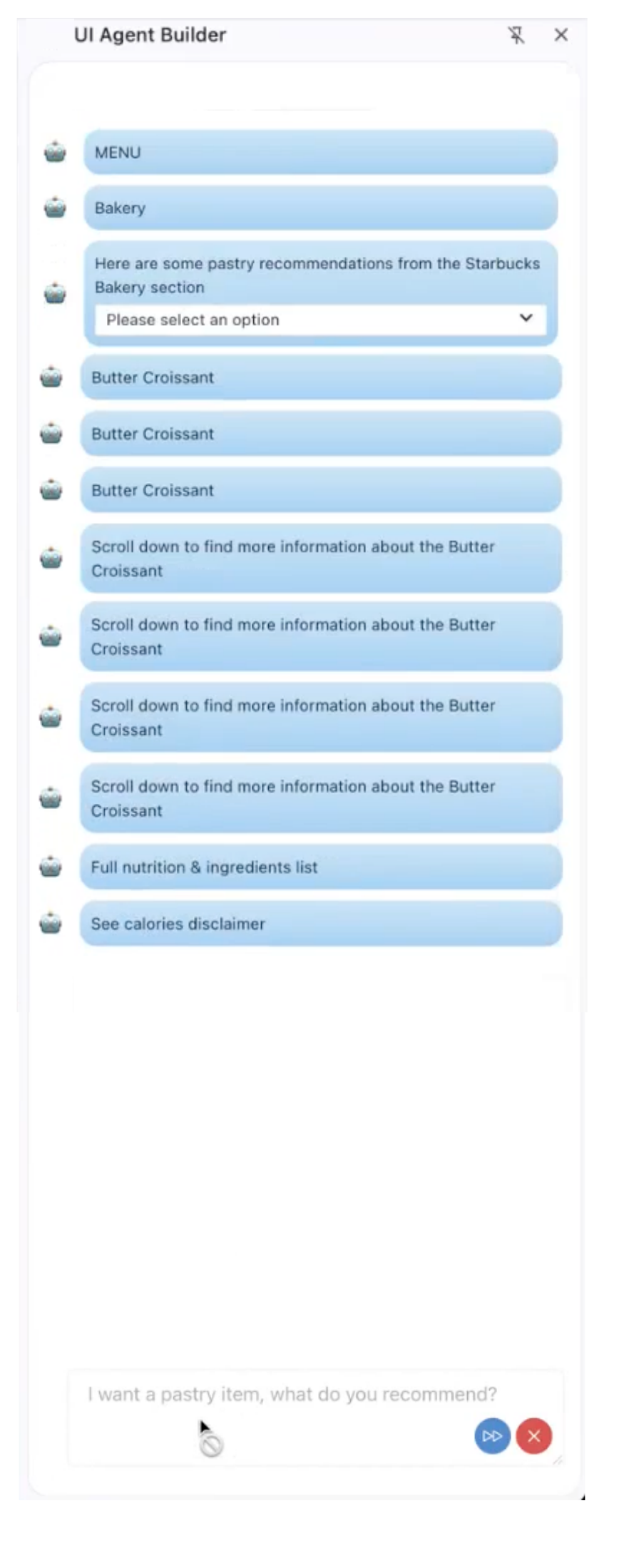} 
\caption{
The agent UX prototype that P6 created during the \emph{in situ} agent UX prototyping study for the input \pQuote{I want a pastry item, what do you recommend?}{6}
}
\label{fig:p6-example-prototype}
\end{figure}

\subsubsection{P6 agent UX prototypes}
The agent UX prototype created by P6 is shown in Figure~\ref{fig:p6-example-prototype} with the user query \pQuote{I want a pastry item, what do you recommend?}{6} to order a butter croissant.
This agent was designed to not be interactive with a workflow of \texttt{Start} $\rightarrow$ \texttt{UI Actions} $\rightarrow$ \texttt{End}.
However, P6 added a guideline that described an exception to this, which said while exploring options, the agent should \texttt{show top 3 to 5 items} (P6) in the category.
P6 also chose to show the action descriptions for UI actions but added a guideline to shorten the UI action descriptions.
When this agent was run, the agent provided recommendations in accordance to P6's instructions via a drop-down menu and displayed very brief descriptions of each UI action (e.g., \emph{``MENU''}, \emph{``Bakery''}).

\begin{figure}[t!]
\centering
\includegraphics[trim=0 25 1500 0, clip, width=0.37\linewidth, keepaspectratio, page=5]{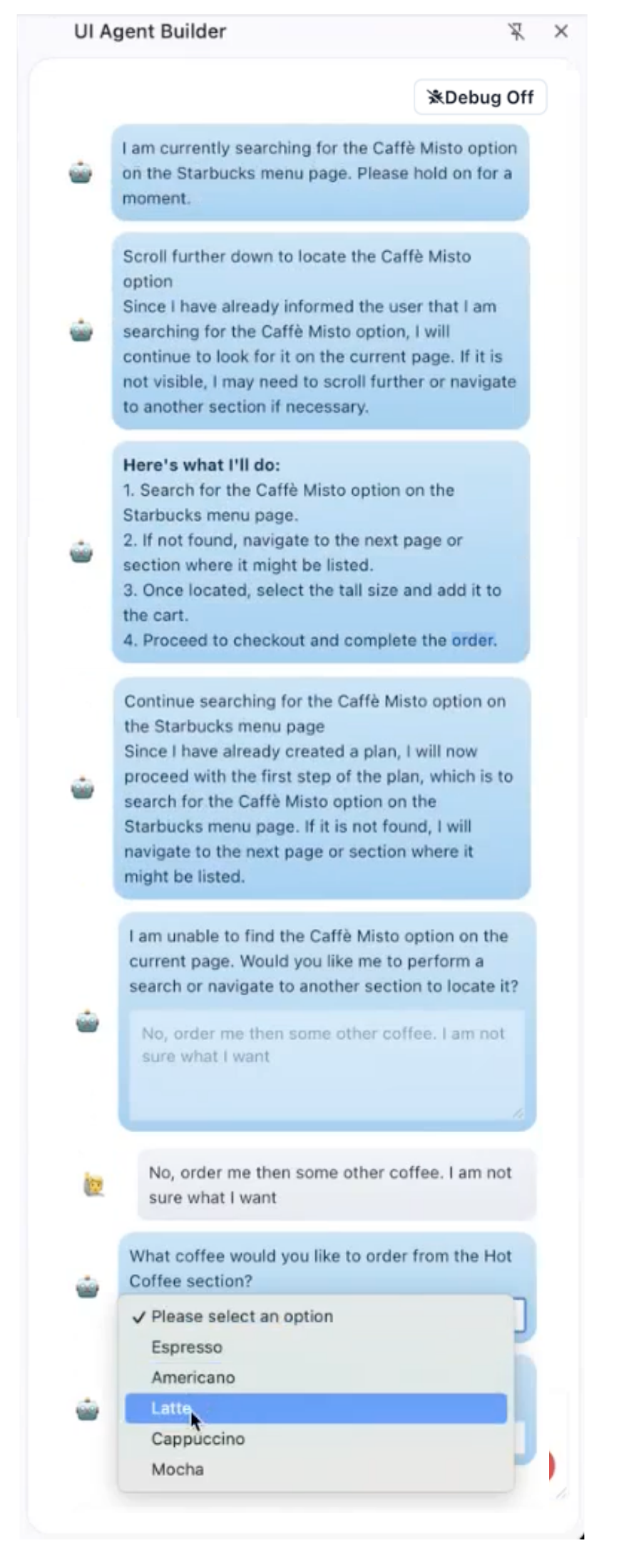} 
\caption{
The agent UX prototype that P14 created during the \emph{in situ} agent UX prototyping study for the input \pQuote{Order me a tall Caffe Misto}{14}.
}
\label{fig:p14-example-prototype}
\end{figure}

\subsubsection{P14 agent UX prototypes}
One of the agent UX prototypes created by P14 is shown in Figure~\ref{fig:p14-example-prototype}, where P14 asked, \pQuote{Order me a tall Caffe Misto}{14}.
P14 designed this agent to show more information about UI actions (i.e., action description, agent reasoning).
P14 designed the agent's workflow to handle agent errors, wherein upon encountering an error, the agent would show a plan that the user would confirm the next action. 
If the user rejected the suggestions, the agent would further interact with the user (i.e., \texttt{Start} $\rightarrow$ ... $\rightarrow$ \texttt{UI Actions} $\rightarrow_{agent\_error}$ \texttt{Plan} $\rightarrow$ \texttt{Confirm} $\rightarrow_{confirmation\_declined}$ \texttt{Interact} $\rightarrow$ \texttt{End}).
While running this agent on the user query, P14's agent was unable to find the item. 
It therefore followed the workflow that P14 defined: it displayed a plan, confirmed with the user about the next step of actions (\emph{``Would you like me to perform a search or navigate to another section to locate [the item]?''}), which P14 declined (\pQuote{No, order me some other coffee...}{14}).
The agent followed up by asking what other coffee the user wanted.

\subsection{Agent UX Prototyping Exercise Instructions}
\label{sec:in-situ-study-instructions}

\subsubsection*{\textbf{Description}}
You are a business owner looking to have an UI agent help make customer orders on the [popular coffee shop chain's] website. Design an agent that achieves this task for your customers so they have a positive user experience, while considering the different kind of user inputs that might occur.

More specifically, \underline{design a system prompt} to tell the agent which actions to use and when to provide the best user experience. You should also decide \underline{what information the agent should reveal in the chat and UI} about its actions. You can assume the user's credit card information is linked to the [popular coffee shop chain's] website.

\textbf{Important:} The agent may not be able to complete the actions on its own. It is more important to achieve a good user experience with the agent, rather than getting the agent to perform actions correctly and independently.

\subsubsection*{\textbf{User Inputs}}
You will start the interaction from the [populare coffee shop chain's] homepage. Here are some situations that you may encounter that your agent should handle:

\begin{itemize}
    \item "Order me a coffee. I’m not sure what I want."
    \item "Order me a tall Caffè Misto."
    \item "Order me a tall iced chai latte and a butter croissant."
    \item "Get me a croissant." (to order a chocolate croissant)
    \item "I’m not sure what I want to order. Give me a couple of ideas based on what I’ve ordered in the past."
\end{itemize}

\subsubsection*{\textbf{User Background}}
The agent already knows some information about the user, including:

\begin{itemize}
    \item The user picks up their order at 442 Terry Ave N, Seattle, WA 98109, known as the Terry \& Republican store
    \item Their password to the [popular coffee shop chain's] account is "{\textbf{REDACTED}}2025"
    \item The user’s previous orders include a birthday cake pop, Caffè Misto, and a cold brew at the [popular coffee shop chain]
\end{itemize}

\subsubsection*{\textbf{Agent Guide}}
Here are the actions that the agent can do and how they look like. You may need to adjust your prompt or the agent's appearance to tell the prompt when to use the actions. \emph{[Show images of each agent tool from the tutorial slides]}

\end{document}